\documentclass[twocolumn,showpacs,preprintnumbers,amsmath,amssymb]{revtex4}

\usepackage{graphicx}
\usepackage{dcolumn}
\usepackage{bm}

\begin{document}
\preprint{APS/123-QED}

\title{Defect generation and deconfinement on corrugated topographies}

\author{Vincenzo Vitelli}
\author{David R. Nelson}%
\affiliation{%
Department of Physics, Harvard University, Cambridge MA, 02138
}%

\date{\today}

\begin{abstract}

We investigate topography-driven generation of defects in liquid
crystals films coating frozen surfaces of spatially varying
Gaussian curvature whose topology does not automatically require
defects in the ground state. We study in particular
disclination-unbinding transitions with increasing aspect ratio
for a surface shaped as a Gaussian bump with an hexatic phase
draped over it. The instability of a smooth ground state texture
to the generation of a $single$ defect is also discussed. Free
boundary conditions for a single bump are considered as well as
periodic arrays of bumps. Finally, we argue that defects on a bump
encircled by an aligning wall undergo sharp deconfinement
transitions as the aspect ratio of the surface is lowered.

\end{abstract}

\pacs{Valid PACS appear here}
\maketitle

\section{\label{sec:intr}Introduction}

The melting of a two dimensional crystal can occur continuously
via two second order topological phase transitions characterized
by the successive unbinding of dislocation and disclination pairs.
At low temperatures, dislocations are suppressed due to their
large energy cost, but as the temperature is increased, the
entropy gained by creating defects overcomes their cost in elastic
energy and dislocation unbinding occurs to reduces the overall
free energy of the system \cite{nels79, kost73, youn79}. The
quasi-long range order of the crystal is thus destroyed leading to
an hexatic phase that still preserves quasi-long range
orientational order. This phase can be characterized by a complex
order parameter with six-fold symmetry. As the temperature is
increased still further, an additional disclination-unbinding
transition occurs and the hexatic order is finally lost in an
isotropic liquid phase \cite{nels79}.

Experimental evidence for hexatic order and defect-mediated
melting has been obtained in systems as diverse as free standing
liquid crystal films \cite{chou98}, Langmuir-Blodgett surfactant
monolayers \cite{knob92}, two-dimensional magnetic bubble arrays
\cite{sesh91}, electrons trapped on the surface of liquid helium
\cite{grim79, glat88, devi84}, two-dimensional colloidal crystals
\cite{murr92, zahn99} and self-assembled block copolymers
\cite{sega03}.

The unbinding of defects in the plane is entropically driven and
at low temperature defects are tightly bound. By contrast, on
surfaces with non zero (integrated) Gaussian curvature, excess
defects $must$ be present even at very low temperatures. The
theory of topological defects in ordered phases confined to frozen
topographies with positive or negative Gaussian curvature has been
investigated previously; see, e.g., \cite{sach84,Nels83,Nelsbook}.
As a general rule, regions of positive or negative curvature
(valleys, hills or saddles) lead to unpaired disclinations in the
ground state, possibly screened by clouds of dislocations. These
clouds can in turn condense into grain boundaries at low
temperature. The predictions of recent studies of crystalline
order on a sphere \cite{Bowi98} have been confirmed in elegant
studies of colloidal particles packed on the surface of water
droplets in oil \cite{Baus03}. Investigations of the physics of
defects in curved spaces have also been carried out for
fluctuating geometries \cite{Davi87, Nels87, guit90, Park96}. The
dynamics of hexatic order on fluctuating spherical interfaces was
studied in Ref. \cite{Lenz03}. Quenched $random$ topographies in
the limit of small deviations from flatness were investigated in
Ref. \cite{sach84} .

In the present work, we investigate topography-driven generation
of defects on simple frozen surfaces with spatially varying
Gaussian curvature whose topology does $not$ automatically enforce
their presence in the ground state. We study in particular a
two-dimensional "bump" with a Gaussian shape and dimension large
compared to the particle spacing. For such a hilly landscape, flat
at infinity, the $geometric$ control parameter is an aspect ratio
given by the bump height divided by its spatial extent. Consider a
hexatic phase draped over such a bump. For small bumps, the ideal
hexatic texture is distorted, but there are no defects in the
ground state. As the aspect ratio is increased, we find that
disclination pairs progressively unbind at $T=0$ in a sequence of
transitions occurring at critical values of the aspect ratio. The
defects subsequently position themselves to partially screen the
Gaussian curvature. For bumps embedded in surfaces of sufficiently
small spatial extent, a second instability of the smooth ground
state needs to be considered. In this case, the energy stored in
the field can be lowered by generating a $single$ positive defect
at the center of the bump. Novel effects also arise when the hilly
surface is encircled by an aligning circular wall that insures a
$2\pi$ rotation of the orientational order in the ground state. In
this case, some of the positive defects required to match the
curvature of the boundary are confined to a hemispherical cup
centered on the bump, provided the aspect ratio $\alpha$ is larger
than a critical value $\alpha_{D}$. When $\alpha$ is lowered below
$\alpha_{D}$, the positive defects originally "trapped" in the
hemispherical cup start undergoing a series of sharp
"deconfinement transitions", as they progressively migrate to new
equilibrium positions dictated by boundary conditions and the
finite system size. We also suggest possible ground states for
periodic arrays of bumps, like those on the bottom of an egg
carton.

A natural arena to experimentally study the interplay between
geometry and defects is provided by thin copolymer films on
SiO$_{2}$ patterned substrates \cite{Kram}. Flat space experiments
by Segalman $et$ $al.$ have already demonstrated that spherical
domains in block copolymer films form hexatic phases
\cite{sega03}.

Our results for hexatics on frozen topographies also apply to
other XY-like models, as might be appropriate for tilted
surface-active molecules on curved substrates with interactions
which favor alignment. The results are relevant as well to
two-fold nematic order on frozen topographies. In both cases, we
expect qualitatively similar defect unbinding transitions,
although the equivalence becomes more exact in the one Frank
constant approximation \cite{deGennesbook}. Related results have
been obtained recently for order on a torus \cite{Bowi03}. Even
though the $integrated$ Gaussian curvature vanishes, defects
appear in the ground state in the limit of fat torii, unless the
number of degrees of freedom is very large.

The outline of this paper is as follows. In Section II, the
relevant mathematical formalism is introduced and used to
highlight similarities and differences between defects on surfaces
of varying curvature and electrostatic charges in a non-uniform
background charge distribution in flat space. As an example, we
calculate the distorted, but defect-free, ground state texture of
a hexatic confined to a surface shaped as a "Gaussian bump" for
aspect ratios below the first disclination-unbinding instability.
In Section III, we investigate curvature-induced defect formation
for an isolated bump and a periodic array of bumps. In section IV,
defect deconfinement is discussed and in section V various
experimental issues related to our analysis are highlighted along
with some directions for future work. The development of the
mathematical formalism is largely relegated to Appendices. In
Appendix \ref{appA} the Green's function for the covariant
Laplacian is derived by means of conformal transformations. In
Appendix B, we introduce a geometric potential whose source is the
Gaussian curvature. In Appendix \ref{appC}, we present the general
formula for the energy of textures with defects in terms of the
two functions derived in Appendix \ref{appA} and \ref{appB}. We
thus explore the existence of position-dependent defect
self-interactions that arise from the varying Gaussian curvature.
Finally, boundary effects are discussed in Appendix \ref{appD}.

\section{\label{Hexa}Hexatic order on a surface}

\subsection{\label{elec}Electrostatic analogy}

The free energy for hexatic degrees of freedom embedded in an
arbitrary frozen surface can be written as
\begin{equation}
F=\frac{K_{A}}{2}\int dA \!\!\!\!\! \quad D_{\alpha}n^{\beta}({\bf
u})D^{\alpha}n_{\beta}({\bf u}) \!\!\!\! \quad , \label{free
energy}
\end{equation}
where ${\bf u}=\{u_{1},u_{2}\}$ is a set of internal coordinates,
${\bf n}({\bf u})$ is a unit vector in the tangent plane,
$D_{\alpha}$ is the covariant derivative with respect to the
metric of the surface and $dA$ is the infinitesimal surface area
\cite{Nels87,Davi87,Park96,Davidreview}. The generalization to
systems with a p-fold symmetry is straightforward provided that
the one Frank constant approximation is used and the consequences
of the uniaxial coupling neglected \cite{deGennesbook}. This
choice of free energy implies that the minimal energy
configuration will be given locally by neighboring ${\bf n}({\bf
u})$ vectors which differ only by parallel transport. The
curvature of the surface induces "frustration" in the texture. In
fact, by Gauss' "Theorema egregium" \cite{Struik-book,Kami02},
tangent vectors parallel transported along a closed loop are
rotated by an amount equal to the Gaussian curvature integrated
over the enclosed area. On a sphere, for example, the hexatic
ground state always has twelve excess disclinations as a result of
this frustration \cite{Nels83,lube92}. More generally, the sum of
the topological charges on any closed surface is equal to the
integrated Gaussian curvature.

By introducing a local bond-angle field $\theta({\bf u})$,
corresponding to the angle between ${\bf n}({\bf u})$ and an
arbitrary local reference frame, we can rewrite the hexatic free
energy introduced in Eq.(\ref{free energy}) as:
\begin{equation}
F = \frac{1}{2}K_{A}\int dA \!\!\!\!\! \quad g^{\alpha\beta}
(\partial_{\alpha}\theta - A_{\alpha})(\partial_{\beta}\theta -
A_{\beta})\!\!\!\! \quad , \label{eq:patic-ener}
\end{equation}
where $dA=d^{2}u\sqrt{g}$, $g$ is the determinant of the metric
tensor $g_{\alpha\beta}$ and $A_{\beta}$ is the spin-connection
whose curl is the Gaussian curvature $G({\bf u})$
\cite{Davidreview,Kami02}. The spin connection can be viewed as a
"geometric vector potential". A free energy like Eq.(2) also
describes the charged Cooper pairs implicit in the London theory
of a superconductor well below $T_{C}$. In the superconductor
analogy, the Gaussian curvature plays the role of a (spatially
varying) external magnetic field. For the problem considered here,
however, there are interesting new nonlinear effects associated
with spatial variations in the metric.

A detailed analysis of the free energy of Eq.(\ref{eq:patic-ener})
for a bumpy surface with free and circular boundary conditions is
presented in Appendices (\ref{appC}) and (\ref{appD}). Here we
only sketch the main steps and conclusions. The free energy can be
readily converted into a Coulomb gas model by using the relation
\begin{equation}
\gamma^{\alpha\beta}\partial_{\alpha}(\partial_{\beta}\theta-A_{\beta})
= s({\bf u})- G({\bf u}) \equiv n({\bf u}) \!\!\!\! \quad ,
\label{eq:curl-vel}
\end{equation}
where $\gamma^{\alpha\beta}$ is the covariant antisymmetric
tensor, $G({\bf u})$ is the Gaussian curvature and $s({\bf
u})\equiv\frac{1}{\sqrt{g}}\sum_{i=1}^{N_{d}}q_{i}\delta({\bf
u}-{\bf u}_{i})$ is the disclination density with $N_d$ defects of
charge $q_{i}$ at positions ${\bf u}_{i}$. The final result is an
effective free energy whose basic degrees of freedom are the
defects themselves \cite{Park96,Bowi98}:
\begin{eqnarray}
F &=& \frac{K_{A}}{2}\int\! dA \!\!\!\!\! \quad \int \! dA'
\!\!\!\quad  n({\bf u}) \!\!\!\quad \Gamma({\bf u},{\bf u}')
\!\!\!\quad n({\bf u}') \!\!\!\! \quad , \label{Coulomb
model-energy}
\end{eqnarray}
where $n({\bf u})$ is defined in Eq.(\ref{eq:curl-vel}). The
Green's function $\Gamma({\bf u},{\bf u}')$ is calculated (see
Appendix \ref{appA}) by inverting the Laplacian defined on the
surface
\begin{eqnarray}
\Gamma({\bf u},{\bf u}')\equiv -
\left(\frac{1}{\Delta}\right)_{{\bf u}{\bf u}'} \!\!\!\! \quad ,
\label{invert}
\end{eqnarray}
and we have suppressed for now defect core energy contributions
which reflect the physics at microscopic length scales.
Eq.(\ref{Coulomb model-energy}) can be understood by analogy to
two dimensional electrostatics, with the Gaussian curvature
$G({\bf u})$ (with sign reversed) playing the role of a
non-uniform background charge distribution and the topological
defects appearing as point-like sources with electrostatic charges
equal to their topological charge $q_{i}$. As a result, the
defects tend to position themselves so that the Gaussian curvature
is screened: the positive ones on peaks and valleys and the
negative ones on the saddles of the surface. However, this analogy
does neglect position-dependent self-interactions
\cite{Vite-Turn04}, but since these are quadratic in the charge
they are negligible for hexatics. Hence positive disclinations of
minimal topological charge $q_{i}=\frac{2 \pi}{6}$ continue to be
attracted to positive curvature (see Appendix \ref{appC}).

\begin{figure*}
\includegraphics[width=0.85\textwidth]{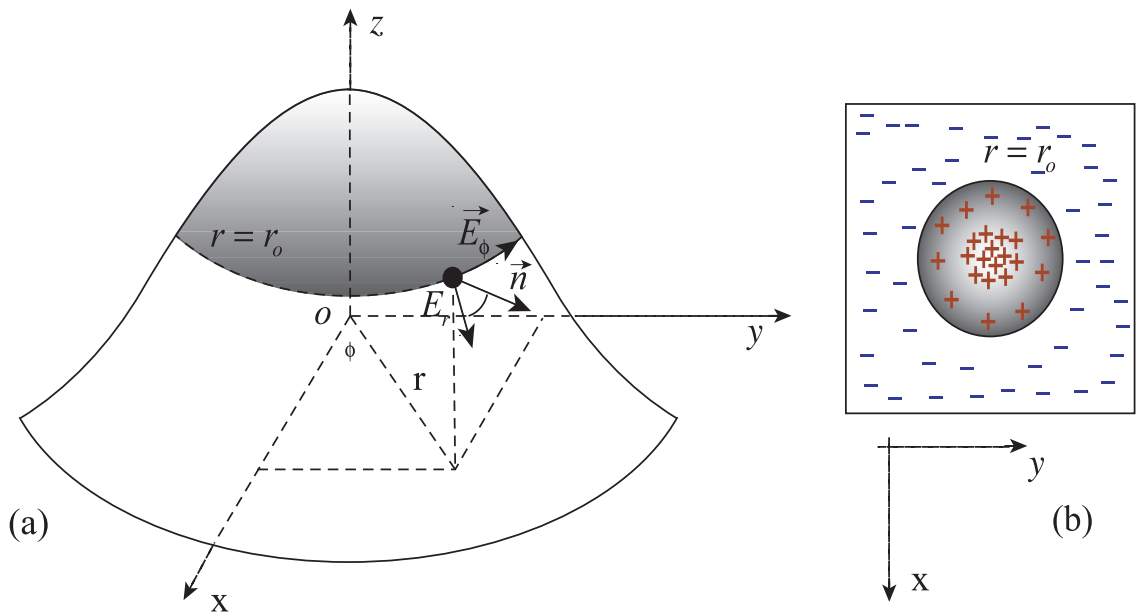}
\caption{\label{bump}(a) The vector field ${\bf n}$ is confined to
a surface shaped as a Gaussian. (b) Top view of (a) showing a
schematic representation of the positive and negative Gaussian
curvature as a background "charge" distribution that switches sign
at $r=r_{0}$. Note that, according to the electrostatic analogy, a
positive (negative) distribution of Gaussian curvature corresponds
to negative (positive) topological charge density.}
\end{figure*}

More generally, we can consider p-fold symmetric order parameters
with minimum charge defects $\pm \frac{2 \pi}{p}$. The case $p=1$
corresponds to tilt order of absorbed molecules and $p=2$
describes $2D$ nematics. The cases $p=4$ and $p=6$ describe
tetradic and hexatic phases respectively \cite{lube92}. Strictly
speaking, Eq.(1) only describes the cases $p=1$ and $p=2$ in the
one-Frank-constant approximation \cite{deGennesbook}. Most of our
discussion focuses on topography-driven transitions on a model
surface shaped like a bell curve or "Gaussian bump" (see Fig.
\ref{bump}),
but the same mathematical approach can be readily carried over to
study arbitrary surfaces of revolution that are topologically
equivalent to the plane. Furthermore, we do not expect the results
of this analysis to depend qualitatively on the azimuthal symmetry
of the surface, which is assumed purely for reasons of
mathematical convenience.

\begin{figure}
\includegraphics[width=0.47\textwidth]{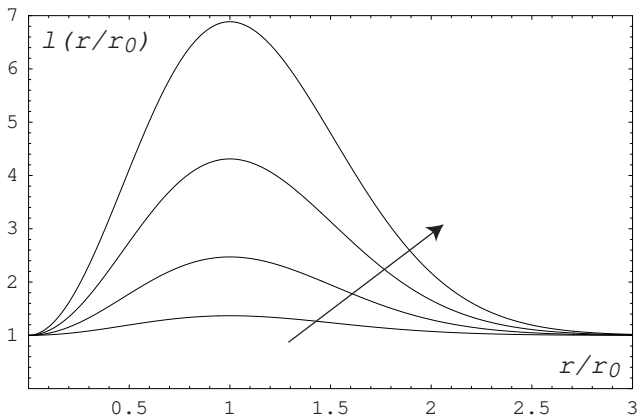}
\caption{\label{fig:l(r)} Plot of $l(\frac{r}{r_{0}})$ as a
function of the dimensionless radial coordinate $\frac{r}{r_{0}}$
for $\alpha=1,2,3,4$. The arrow is oriented in the direction of
increasing $\alpha$.}
\end{figure}

Points on our model surface embedded in three dimensional
Euclidean space are specified by a three dimensional vector ${\bf
R}(r,\phi)$ given by
\begin{equation}{\bf R}(r,\phi)=\left(\begin{array}{c} r\cos\phi \\ r\sin\phi \\
h \exp\left(-\frac{r^2}{2r_0^{2}}\right)\end{array}\right)
\!\!\!\! \quad , \label{eq-coord-polar}
\end{equation}
where $r$ and $\phi$ are plane polar coordinates in the $x$$y$
plane of Fig. \ref{bump}. It is useful to characterize the
deviation of the bump from a plane in terms of a dimensionless
aspect ratio
\begin{equation}
\alpha \equiv \frac{h}{r_{0}} \!\!\!\! \quad. \label{aspect ratio}
\end{equation}
The two orthogonal tangent vectors ${\bf t}_{r}\equiv
\frac{\partial {\bf R}}{\partial r}$ and ${\bf t}_{\phi} \equiv
\frac{\partial {\bf R}}{\partial \phi}$ can be normalized to
define the Vierbein (orthonormal basis vectors) ${\bf E}_{r}$ and
${\bf E}_{\phi}$ respectively. The components of the spin
connection introduced in Eq.(\ref{eq:patic-ener}) are given by
${\bf A_{\alpha}} = {\bf E}_{r} \cdot \partial_{\alpha} {\bf
E}_{\phi}$ \cite{Davidreview,Kami02}. This leads to a vanishing
radial component $A_r$ and
\begin{equation}
A_{\phi}=-\frac{1}{\sqrt{l(r)}} \!\!\!\! \quad , \label{A_phi}
\end{equation}
where the important $\alpha$-dependent function $l(r)$ (see Fig.
\ref{fig:l(r)}) is defined by
\begin{equation}
l(r) \equiv 1+ \frac{{\alpha}^2 r^2}{r_{0}^2} \exp
\left(-\frac{r^2}{r_{0}^2}\right) \!\!\!\! \quad , \label{l}
\end{equation}
and it is equal to the radial component of the diagonal metric
tensor, $g_{\alpha\beta}$,
\begin{equation}
g_{\alpha\beta} = \begin{pmatrix} l(r) & 0 \\
  0 & r^2
\end{pmatrix} \!\!\!\! \quad .
\label{metric-mat}
\end{equation}
Note that the $g_{\phi\phi}$ entry is equal to the flat space
result $r^{2}$ in polar coordinates while $g_{rr}=l(r)$ is
modified in a way that depends on $\alpha$ but tends to the plane
result $g_{rr}=1$ for small and large $r$, as illustrated in Fig.
\ref{fig:l(r)}.

The Gaussian curvature for the bump is readily found from the
eigenvalues of the second fundamental form \cite{Dubrovinbook},
\begin{equation}
G_{\alpha}(r)=\frac{\alpha^2 e^{-\frac{r^2}{r_{0}^2}}}{r_{0}^2
\!\!\!\! \quad l(r)^2} \left(1-\frac{r^2}{r_{0}^2}\right) \!\!\!\!
\quad . \label{Gaussian curvature}
\end{equation}
Note that $\alpha$ controls the order of magnitude of $G(r)$ and
that $G(r)$ changes sign at $r=r_{0}$ (see Fig. \ref{bump}b). The
integrated Gaussian curvature $\Delta G(r)$ inside a cup of radius
$r$ centered on the bump is
\begin{equation}
\Delta G(r) = 2 \pi \left(1- \frac{1}{\sqrt{l(r)}}\right) \!\!\!\!
\quad , \label{intgr Gauss curv}
\end{equation}
which vanishes as $r \rightarrow \infty$. Eq.(\ref{intgr Gauss
curv}) also shows that the positive Gaussian curvature enclosed
within the radius $r_{0}$ (see Fig. \ref{bump}) approaches $2\pi$
for $\alpha\gg1$, half the integrated Gaussian curvature of a
sphere.

\subsection {\label{defe}Defect free texture}

For small values of the aspect ratio $\alpha$, the minimal energy
texture for the hexatic will be free of defects. The ground state
configuration $\theta_{o}({\bf u})$ satisfies the differential
equation
\begin{equation}
D_{\alpha}D^{\alpha}\theta_{0}- D^{\alpha}A_{\alpha}=0 \!\!\!\!
\quad ,
\label{eq-motion}
\end{equation}
which results from minimizing the free energy in
Eq.(\ref{eq:patic-ener}) with respect to the field $\theta(\bf u)$
for fixed $A_{\alpha}$. When expressed in terms of the coordinates
in Eq.(\ref{eq-coord-polar}), the solution of Eq.(\ref{eq-motion})
reads:
\begin{equation} \theta_{o}({\bf u})= - \phi + c  \!\!\!\! \quad ,
\label{ground state angle}
\end{equation}
where $c$ is an arbitrary constant. The smooth ground state
texture is thus obtained if the director ${\bf n}$ forms an angle
$\theta_{o}({\bf u})= - \phi + c$ with respect to the spatially
varying basis vector ${\bf E}_{r}$. Note that a solution of the
form $\theta_{o}({\bf u})=c$ represents a defect of charge $q=
2\pi$ in this "rotating" system of coordinates.

\begin{figure*}
\includegraphics[width=0.85\textwidth]{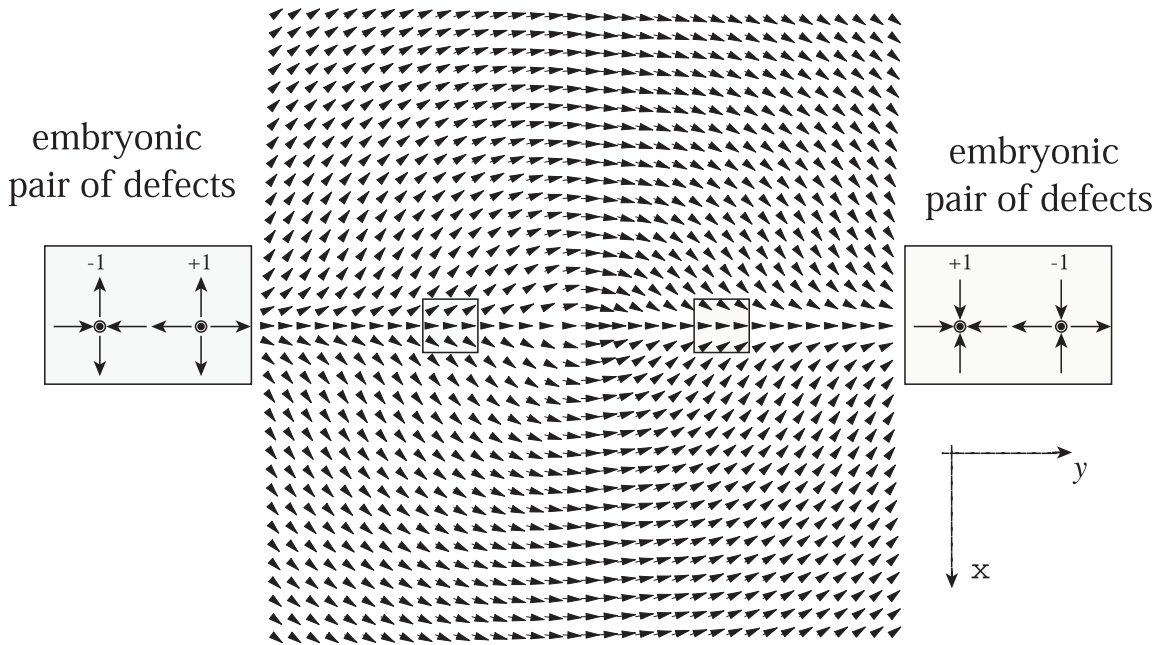}
\caption{\label{defect free texture}Projected ground state texture
for an XY model on the bump, with the boundary condition that the
vector field is parallel to the $y$-axis at infinity. The two
insets show the defect pairs suggested by two regions of large
frustration, which lie close to a circle of radius $r_{0}$.}
\end{figure*}

As an illustration, consider the projection on the plane of the
minimal energy texture of an XY model ($p=1$) as shown in Fig.
\ref{defect free texture}.
The arrows represent the orientation of tilted molecules on this
surface in the one Frank constant approximation. The field clearly
displays strong frustration along a direction determined by the
choice of the constant $c$ in Eq.(\ref{ground state angle}). If
the bump is positioned within two very distant walls parallel to
the y-axis which impose tangential boundary conditions on the
molecular tilts, the "preferred" direction will be along $\hat{y}$
\footnote{In case $distant$ walls are present, the free solution
of Eq.(\ref{ground state angle}) is slightly modified to account
for the new boundary conditions. This is accomplished by the
method of images or conformal transformations \cite{Smirnov3.2}.}.
The texture displayed in Fig. \ref{defect free texture} can be
interpreted as resulting from embryonic pairs of defect dipoles
along the line $x=0$. The distortion energy $F_{0}$ of this ground
state is given by:
\begin{equation}
F_{0} = \frac{1}{2}K_{A} \int dA \!\!\!\!\! \quad g^{\alpha\beta}
(\partial_{\alpha}\theta_o - A_{\alpha})(\partial_{\beta}\theta_o
- A_{\beta})\!\!\!\! \quad . \label{Fo-I}
\end{equation}
This expression can be evaluated for an infinitely large system by
using Eq.(\ref{ground state angle}) and the explicit form of the
spin connection derived in Section \ref{elec}, with the result:
\begin{equation}
F_{0} = \pi K_{A} \int_0^{\infty} dr \frac{\left(1-\sqrt{l(r)}
\!\!\!\! \quad \right)^{2}}{r \sqrt{l(r)}} \!\!\!\! \quad .
\label{Fo-II}
\end{equation}
It follows from Eq.(\ref{Fo-II}) that the ground state energy is a
monotonically increasing function of the aspect ratio,
proportional to ${\alpha}^{4}$ for small $\alpha$. As we shall
see, for large enough $\alpha$, it can be energetically preferable
to reduce this energy by introducing defect pairs into the
texture. It is convenient to rewrite Eq.(\ref{Fo-I}) in terms of
the Gaussian curvature $G(r)$ and the Green function $\Gamma({\bf
u}, {\bf u'})$ discussed in Appendix \ref{appA} \cite{Davidreview}
\begin{equation} F_{0}=\frac{K_{A}}{2} \int dA \!\!\!\!\! \quad \int \! dA' \!\!\!\quad G({\bf u}) \!\!\!\quad \Gamma({\bf u},{\bf u}') \!\!\!\quad G({\bf u}') \!\!\!\! \quad
.
\label{free texture energy}
\end{equation}
This result is what one obtains by setting all $q_i=0$ in
Eq.(\ref{Coulomb model-energy}). The details of the mathematical
derivation are relegated to Appendix \ref{appC}.

Although this result correctly represents the zero temperature
limit of the vector model, corrections may be appropriate to
describe the physics of ordered phases at finite temperature.
"Spin-wave" excitations (i.e., quadratic fluctuations of the order
parameter about the ground state texture) can be accounted for by
integrating out the longitudinal fluctuations $\theta'({\bf u})$
around the ground state configuration $\theta_{o}({\bf u})$. By
letting $\theta=\theta_{0}+\theta'$ in Eq.(\ref{eq:patic-ener})
and using Eq.(\ref{eq:curl-vel}) we obtain \cite{Davi87,Park96}:
\begin{equation}
F  = F_{0} + \frac{1}{2}K_{A} \int dA \!\!\!\!\! \quad
g^{\alpha\beta}
\partial_{\alpha}\theta'\partial_{\beta}\theta'\!\!\!\! \quad ,
\label{eq:spin-wave}
\end{equation}
The longitudinal variable $\theta'({\bf u})$ appears only
quadratically in $F$ and the trace over $\theta'({\bf u})$ can be
explicitly performed with the result \cite{Poly81}:
\begin{equation}
\int {\cal D}\theta' e^{-\frac{\beta K_{A}}{2} \int dA \!\!\!\!\!
\quad g^{\alpha\beta}
\partial_{\alpha}\theta'\partial_{\beta}\theta'} = e^{-\beta
F_{L}}\!\!\!\! \quad ,
\end{equation}
where $F_{L}$ is the Liouville action,
\begin{equation}
\beta F_{L} = c \int dA \!\!\!\!\! \quad - \frac{K_{A}}{24} \int
dA \!\!\!\!\! \quad \int \! dA' \!\!\!\quad G({\bf u}) \!\!\!\quad
\Gamma({\bf u},{\bf u}') \!\!\!\quad G({\bf u}') \!\!\!\! \quad .
\end{equation}
The first term in this expression is a constant proportional to
the fixed surface area of the frozen topography and will be
suppressed in what follows. The remaining term causes a shift in
the coupling constant appearing in Eq.(\ref{free texture energy})
from $K_{A}$ to $K_{A}'= K_{A} - \frac{k_{B} T}{12 \pi}$
\cite{Davi87,Park96}. This "entropic" correction to the coupling
constant $K_{A}$ at finite temperature also arises when defects
are present.

The energy in Eq.(\ref{free texture energy}) represents an
intrinsic, irreducible energy cost of geometric frustration for
textures without defects. As we shall see, defects can reduce this
frustration. However, for small values of $\alpha$ the energy cost
of this frustration will still be lower than the core energies
associated with the creation of the unbound defects and the work
necessary to tear them apart.
\subsection{\label{ener}Energetics of defect pairs on a Gaussian bump}
A quantitative understanding of the energetics of defects on a
fixed topography is essential to calculate the critical value(s)
of the aspect ratio above which defect-unbinding becomes
energetically favorable. The first step is to calculate the
Green's function $\Gamma({\bf u},{\bf u'})$ that governs the
"coulombic" interaction among defects and between each defect and
the Gaussian curvature. The inversion of the curved space
Laplacian can be more easily accomplished by employing a set of
"isothermal coordinates", such that the resulting Green's function
reduces to the familiar logarithm of two dimensional
electrostatics. As shown in Appendix \ref{appA} the final result
in terms of the original polar coordinates reads:
\begin{eqnarray} \Gamma({\bf u},{\bf u'})&=& -\frac{1}{4 \pi}
\ln[\Re(r)^{2}+\Re(r')^{2} \nonumber\\
&-& 2\Re(r)\Re(r')\cos (\phi-\phi')] + c \!\!\!\quad ,
\label{green}
\end{eqnarray}
where the function $\Re(r)$ can be thought of as a radial
coordinate in the conformal plane resulting from adopting an
isothermal set of coordinates (see Appendix \ref{appA})
\begin{eqnarray} \Re(r)= r \!\!\!\!\quad e^{-\int_{r}^{\infty}
\frac{dr'}{r'}\left(\sqrt{l(r')}-1\right)} \!\!\!\quad ,
\label{green-coord}
\end{eqnarray}
and $l(r)$ is the $\alpha$-dependent function introduced in
Eq.(\ref{l}). The constant $c$ depends on the physics at short
distances, which is discussed in Appendix \ref{appA}.

The Green function in Eq.(\ref{green}) corresponds to free
boundary conditions at infinity and preserves the cylindrical
symmetry of the metric. It differs from the familiar result in
flat space by a non-linear radial stretch corresponding to a
smooth deformation of the bump into a flat disk. This Green's
function determines an attractive interaction for the defect
dipole pair. However, the Gaussian curvature of the bump also
generates a geometric potential that tries to pull the
disclination dipole apart. This geometric interaction arises by
combining cross terms between $s({\bf u})$ and $G({\bf u})$ in
Eq.(\ref{Coulomb model-energy}) with the position dependent
self-interactions derived in Appendix \ref{appC}. The resulting
interaction $F_{G}$ between defects and the Gaussian curvature
takes the simple form
\begin{equation}
F_{G} = K_{A} \sum_{i=1}^{N_{d}} q_{i}\left(1-\frac{q_{i}}{4
\pi}\right) V({\bf u}_{i}) \!\!\!\! \quad ,
\label{potential-explicit1}
\end{equation}
where the geometrical potential $V({\bf u})$ is defined as
\begin{equation}  V({\bf u}) \equiv - \int\! dA \! \!\!\!\quad  G({\bf u}') \!\!\!\quad
\Gamma({\bf u},{\bf u'})\!\!\!\quad . \label{Coulomb model2}
\end{equation}
The minus sign in front of this geometric potential insures that
defects of topological charge between zero and $4 \pi$ are
attracted by regions of positive Gaussian curvature
\cite{Vite-Turn04}. For defects with a large topological charge of
$q>4\pi$, the sign of the geometric interaction, $F_G$, is
reversed and, and defects of either sign are pushed away from the
bump. This scenario does not affect the geometry-driven defect
formation discussed in this paper that relies on disclinations
whose charge is well below $4 \pi$. However, as a result of the
position-dependent self interactions, $F_{G}$ is no longer
symmetric under the change $q \rightarrow -q$, as one would expect
on the basis of the electrostatic analogy. The effect of this
asymmetry is small for hexatic order but it increase for liquid
crystals with p-fold order parameter, as $p$ decreases (see Ref.
\cite{Vite-Turn04}, and references therein).

Gauss' law generalized to curved surfaces (see Appendix
\ref{appB}) insures that the geometric force experienced by a
defect of charge $q$ with radial coordinate $r$ will be determined
only by the net curvature enclosed in a circle of radius $r$
centered on the top of the bump. The resulting "electric field" is
radial as expected from electrostatics and is proportional to the
gradient of the geometric potential, which for a Gaussian bump
takes the form (see Appendix \ref{appB})
\begin{equation}
V(r)= - \int_{r}^{\infty} \frac{dr'}{r'}
\left(\sqrt{l(r')}-1\right) \!\!\!\! \quad .
\label{potential-explicit}
\end{equation}
For small values of the aspect ratio $\alpha$ the potential in
Eq.(\ref{potential-explicit}) can be approximated by
\begin{equation}
V(r) \approx - \frac{\alpha^{2}e^{-\frac{r^{2}}{r_0^{2}}}}{4}
\!\!\!\! \quad . \label{potential-approx}
\end{equation}
The resulting force is linear for small $r$ (i.e. near the top of
the bump) and decays like $e^{-\frac{r^{2}}{r_0^{2}}}$ for $r \gg
r_0$. As the aspect ratio $\alpha$ increases, the force generated
by the curvature can overcome the attractive force binding the
defect pair which varies logarithmically for short distances. As a
result, oppositely charged defects that were originally tightly
bound can be separated.

\begin{figure}
\includegraphics[width=0.45\textwidth]{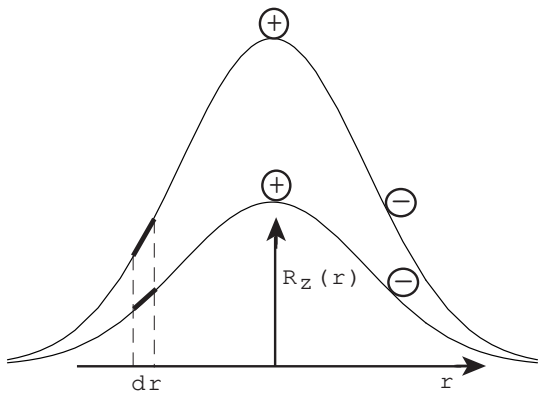}
\caption{\label{greenfunction}Effect of changing the aspect ratio
of the bump on the work needed to pull apart two oppositely
charged defects. Positive and negative defects are represented by
open circles with their sign printed. The line elements
corresponding to the projected length $dr$ for the two aspect
ratios are shown.}
\end{figure}

This argument, however, neglects another complication resulting
from the curvature of the surface: as the aspect ratio is
increased, the Green's function $\Gamma({\bf u},{\bf u}')$ in
Eq.(\ref{green}) and hence the force binding the defects together
$also$ increases. We illustrate this point in Fig.
\ref{greenfunction} for the special case of a positive defect
pinned right on top of the bump and a negative one free to move
downhill at $r$. Upon invoking Equations (\ref{Coulomb
model-energy}) and (\ref{green}), the potential binding the pair,
$V_{pair} \!\!\!\!\! \quad (r)$, can be written down exactly as:
\begin{eqnarray}
V_{pair} \!\!\!\!\! \quad (r) &=& \frac{K_{A} \!\!\!\!\! \quad q^{2}}{2 \pi}\ln\left(\frac{r}{a}\right)+2 q^2 E_{c} \nonumber\\
&-& \frac{K_{A} \!\!\!\!\! \quad q^{2}}{2 \pi} \int_{r}^{\infty}
\frac{dr'}{r'} \left(\sqrt{l(r')}-1 \right) \!\!\!\! \quad .
\label{potential}
\end{eqnarray}
The first term is the flat space Green's function and we have
added two disclination core energies. The last term represents a
"curvature correction" that has the same functional form of the
geometric potential in Eq.(\ref{potential-explicit}), but it
represents a distinct contribution to the total free energy. As
discussed in Appendix \ref{appB}, the pair potential energy,
$V_{pair} \!\!\!\!\! \quad (r)$, can be understood by applying a
generalized Gauss' Law to the bump to determine a force which is
equal to $-\frac{q^{2}}{2\pi r}$. The potential follows by
integrating this force along the bump with the length element $dr
\sqrt{l(r)}$.
Although the force is independent of the aspect ratio, the length
element grows with $\alpha$ (see Fig. \ref{greenfunction}), which
makes the pair more energetically bound for larger values of
$\alpha$. A careful calculation of these effects (including the
contribution associated with the position-dependent self-energies)
reveals that the geometric force still overcomes the binding
interaction for sufficiently large values of $\alpha$.

An estimate of the critical value of $\alpha$ for which the dipole
unbinds can be obtained by comparing the minimal free energy of
the smooth frustrated field arising from Eq.(\ref{free texture
energy}) with the free energy in the presence of defects. The
latter, as follows from Eq.(\ref{Coulomb model-energy}), is
composed of three contributions: the interactions among the
defects, the interaction between the defects and curvature as
given by Eq.(\ref{potential-explicit1}) and the Gaussian curvature
self-interaction. The latter is equal to the minimal free energy
of the smooth frustrated field and is renormalized at finite
temperature in the same way. (Associated with the cutoff is a
microscopic core energy, $E_c$, that we expect to be $independent$
of the defect position on the bump, as long as the radius of
curvature is much greater than any microscopic length scale.) The
remaining two contributions are not renormalized by thermally
induced spin wave fluctuations \cite{Park96}. As shown in Appendix
\ref{appC}, the difference in free energy of a defected texture
described by Eq.(\ref{Coulomb model-energy}) relative to the
defect-free result Eq.(\ref{free texture energy}) can be written
as:
\begin{eqnarray}
\frac{\Delta F (\alpha)}{K_{A}} = \frac{1}{2}\sum_{i=1}^{N_{d}}
\sum_{j \neq i}^{N_{d}} q_{i} q_{j}
\Gamma_{a}(r_{i},\phi_{i},r_{j},\phi_{j}) \nonumber\\
+ \sum_{i=1}^{N_{d}}q_{i}\left(1-\frac{q_{i}}{4
\pi}\right)V(r_{i})+
\frac{E_{c}}{K_{A}}\sum_{i=1}^{N_{d}}q_{i}^{2} \!\!\!\! \quad .
\label{delta f}
\end{eqnarray}
where we have assumed overall charge neutrality for the defect
configuration. The subscript in $\Gamma_{a}$ indicates that a
constant microscopic core radius $a$ has been absorbed in the
definition of the Green function so that the argument of the
logarithm in Eq.(\ref{green}) becomes dimensionless, as in
Eq.(\ref{green-cut}),
\begin{eqnarray}
\Gamma_{a}(r_{i},\phi_{i},r_{j},\phi_{j}) = \Gamma
(r_{i},\phi_{i},r_{j},\phi_{j}) + \frac{1}{2 \pi} \ln a \!\!\!\!
\quad . \label{Gammamod}
\end{eqnarray}
This microscopic cutoff $a$ corresponds to a constant core radius
for each defect and it is of the order of the spacing between the
microscopic degrees of freedom. The sum of microscopic core
energies in the fourth term of Eq.(\ref{delta f}) needs to be
fixed phenomenologically or from models that go beyond simple
elasticity theory \cite{Kleman-book}. Note that both
$\Gamma_{a}(r_{i},\phi_{i},r_{j},\phi_{j})$ and $V(r)$ depend on
$\alpha$. If $\alpha$ becomes sufficiently large so that $\Delta
F$ is less than zero, one or more disclination dipoles unbind in
the hexatic phase at a sequence of critical values
$\alpha_{c_{i}}$. The analogous defects in XY-model textures of
tilted liquid crystal molecules would be $+/-$ vortex pairs.

\section{\label{curv-defe}Curvature induced defect generation}
\subsection {\label{onset} Onset of the defect-dipole instability}
If a dipole is created, say, along the line $r=r_{0}$ of zero
Gaussian curvature, the positive disclination will be pulled
towards the center by the positive curvature while the negative
one will be repelled into the region of negative curvature.
\begin{figure}
\includegraphics[width=0.47\textwidth]{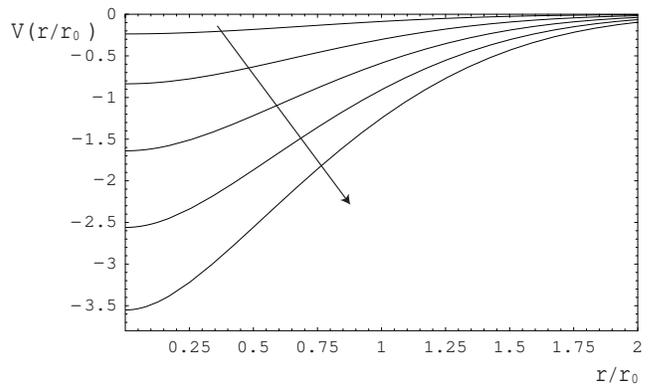}
\caption{Geometric potential $V(r/r_{0})$ as a function of the
dimensionless ratio of $r$ and $r_{0}$ for $\alpha=1,2,3,4,5$.
Note that $ \mid V(r) \mid \ll \!\!\!\!\! \quad \mid V(0) \mid $
for $r \gtrsim r_{0}$. The arrow points in the direction of
increasing $\alpha$.} \label{fig:pot}
\end{figure}
The net result is a reduction of the total free energy of the
order of the depth of the potential well since the logarithmic
binding energy is approximately constant compared to the geometric
potential. An approximate analytical treatment is obtained by
assuming that the positive defect sits right at the center of the
bump and the negative one at a distance of the order of $r_0$. The
validity of this approximation scheme can be checked by
numerically minimizing the energy with respect to the position of
the defects, as discussed in Section (\ref{unbind}). We assume
charge neutrality so that the two defects have equal and opposite
topological charges of magnitude $q$. For order parameters with a
p-fold symmetry, the minimal topological charge is
$q=\frac{2\pi}{p}$. The approximate free energy cost to generate
this defect pair then follows from Equations
(\ref{potential-explicit}),(\ref{potential}) and (\ref{delta f}):
\begin{eqnarray}
\frac{\Delta F (\alpha)}{K_{A}} &\approx& \frac{q^{2}}{2
\pi}\left[\ln\left(\frac{r}{a}\right)+V(r)\right]+ q\left(1- \frac{q}{4\pi}\right)V(0) \nonumber\\
&-&q\left(1+ \frac{q}{4 \pi}\right)V(r)+ 2 q^{2}
\frac{E_{c}}{K_{A}} \!\!\!\! \quad . \label{delta f-1}
\end{eqnarray}
The internal consistency of the formalism can be checked by
investigating the limit $r \rightarrow a$. As the negative defect
approaches the positive one at the center of the bump, we have
$V(a) \thickapprox V(0)$ and the energy tends to the expected flat
space result $2 q^{2} E_{c}$.

\begin{figure}
\includegraphics[width=0.47\textwidth]{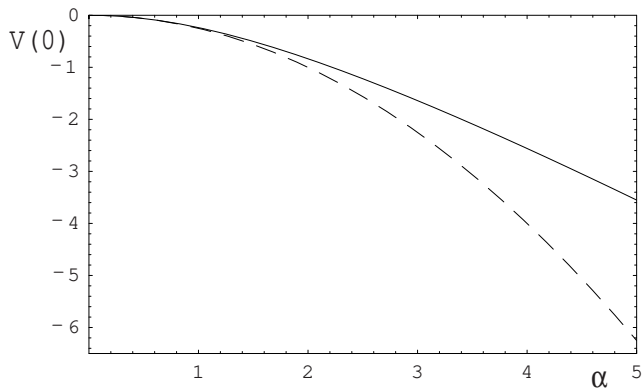}
\caption{Plot of the geometric potential evaluated at the center
of the bump, $V(0)$, for aspect ratios $\alpha$ between 0 and 5.
The continuous line is plotted using the exact form of $V(r)$
while the dashed line is obtained from the low $\alpha$ expansion
given in Eq.(\ref{potential-approx}).} \label{Vo}
\end{figure}

The equilibrium position of the negative defect turns out to be
for $r$ equal to a few $r_0$, so the terms containing $V(r)$ in
Eq.(\ref{delta f-1}) can be dropped as a first approximation
because $V(r)$ decays exponentially (see Fig. \ref{fig:pot}):
\begin{eqnarray}
\frac{\Delta F (\alpha)}{K_{A}} \approx \frac{q^{2}}{2
\pi}\ln\left(\frac{r}{a}\right)+ q\left(1-
\frac{q}{4\pi}\right)V(0) + 2 q^{2} \frac{E_{c}}{K_{A}} \!\!\!\!\!
\quad . \label{delta f-1-bis}
\end{eqnarray}
To estimate the critical value of the aspect ratio for which the
first dipole unbinds, let $r \sim r_0$ and solve for $\alpha_{c}$
in Eq.(\ref{delta f-1-bis}). Because a different choice for the
core energy $E_c$ can be accounted for by rescaling the core size
$a$ in Eq.(\ref{delta f-1-bis}), the condition for unbinding is
\begin{eqnarray}
|V(0)| > \frac{2q}{(4\pi - q)} \ln\left(\frac{r_{0}}{a'}\right)
\!\!\!\!\! \quad , \label{delta f-ap}
\end{eqnarray}
with
\begin{eqnarray}
a'=\!\!\!\!\!\! \quad a \!\!\!\!\! \quad e^{\!\!\!\!\!\! \quad -
\frac{4 \pi E_{c}}{K_{A}}} \!\!\!\!\! \quad . \label{a prime}
\end{eqnarray}
If we know $\frac{E_c}{K_{A}}$ and $\frac{r_{0}}{a}$, the critical
aspect ratio $\alpha_{c}$ can be obtained from inspection of Fig.
\ref{Vo} where $V(0)$ is plotted as a function of $\alpha$.
The Taylor expansion of $V(r)$ derived in
Eq.(\ref{potential-approx}) gives $V(0)\approx -
\frac{{\alpha}^{2}}{4}$ to leading order in $\alpha$. Inspection
of Fig. \ref{Vo} shows that this approximation works sufficiently
well even for aspect ratios of order unity. Upon substituting for
$V(0)$ into Eq.(\ref{delta f-ap}), we obtain an estimate of how
$\alpha_{c}$ depends on $\frac{r_{0}}{a'}$:
\begin{eqnarray}
\alpha_{c}^{2} &\approx& \frac{8q}{(4\pi-q)}
\ln\left(\frac{r_{0}}{a'}\right) \nonumber\\ &\approx&
\frac{8q}{(4\pi-q)} \left[ \ln\left(\frac{r_{0}}{a}\right)+\frac{4
\pi E_{c}}{K_{A}}\right] \!\!\!\!\! \quad . \label{delta f-3}
\end{eqnarray}
Note that a critical height $h_{c}=\alpha_{c} r_{0}$ for defect
unbinding is predicted for fixed $\frac{r_{0}}{a'}$ with defect
charge $q=\frac{2 \pi}{p}$ for all integer values of p. The
validity of this approximate relation is tested in Section
\ref{unbind}.

The continuum theory adopted here is valid in the limit $r_{0}\gg
a$. If $E_{c}$ can be neglected compared to $K_A$, the defect
unbinding instability is triggered when the energy gain derived
from letting the defects screen the Gaussian curvature
(approximately given by $q V(0)$) overcomes the work needed to
pull them apart a distance $r_{0}$. This work, of order
$\frac{q^2}{2 \pi}\ln \left(\frac{r_{0}}{a}\right)$, increases
very slowly with large $\frac{r_{0}}{a}$, hence the continuum
approximation can be satisfied while keeping the work finite. Note
that the result does not depend on the size of the system $R$
because we assume overall disclination charge neutrality and the
assumption that $R\gg r_{0}$. In this limit, boundary effects can
be ignored provided that they do not impose a topological
constraint on the phase of the order parameter. An aligning outer
wall in a circular hexatic sample, for example, would force the
bond angle field to rotate by $2\pi$, leading to six defects in
the ground state even in flat space. The interesting physics which
results is addressed in Section \ref{circ}.

\subsection {\label{unbind} Numerical investigation of defect-unbinding transitions}
\begin{figure}
\includegraphics[width=0.47\textwidth]{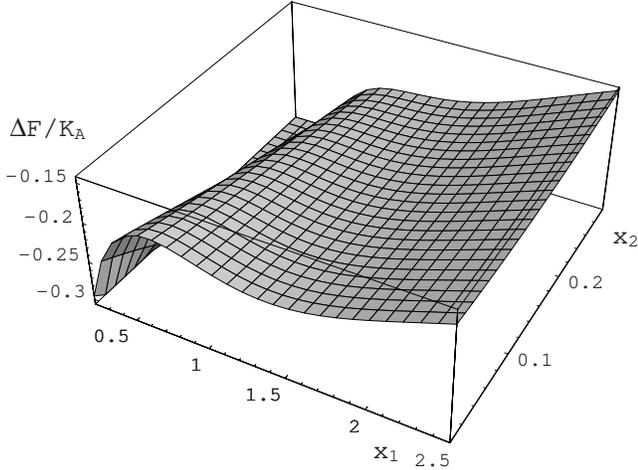}
\caption{Plot of $\Delta F/K_{A}$ versus $x_{1}$ and $x_{2}$ the
positions of the negative and positive defects respectively in
units of $r_{0}$, for $\alpha=1.2$. A constant energy offset equal
to $-\frac{q^2}{2\pi}\ln\left( \frac{r_{0}}{a'}\right)$ has been
neglected. The metastable minimum at $x_{1}\simeq 1.3 \!\!\!\!
\quad r_{0}$ and $x_{2}\simeq 0.2 \!\!\!\! \quad r_{0}$
corresponds to an unbound pair (see Fig. \ref{schematic}a). As
$\alpha$ is decreased further the energy barrier that separates
this minimum from the smooth texture solution (corresponding to
$x_{1}$ approaching $x_{2}$) disappears and the two opposite
defects annihilate.} \label{fig:landscape}
\end{figure}

The disclination unbinding transitions can be investigated more
quantitatively by minimizing numerically $\Delta F (\alpha)$ in
Eq.(\ref{delta f}) with respect to the positions of the defects.
The aspect ratio above which $\Delta F (\alpha)$ becomes negative
corresponds to the threshold value $\alpha_c$ (analogous to a
first order transition) for which the singular field is
energetically favored with respect to the smooth texture of Fig.
\ref{defect free texture}. We emphasize that the energy landscape
can have two minima. The first occurs when two oppositely charged
defects form a closely bound dipole (with separation of the order
of the cutoff $a$) and hence annihilate each other leaving a
smooth texture. The second minimum corresponds to an unbound pair
(with separation of a few $r_{0}$) and it disappears when the
geometric force is too weak to overcome the binding force of the
pair. This scenario occurs for a $finite$ value of $\alpha$
characteristic of the geometry of the substrate  above which the
formation of an unbound dipole is possible, albeit energetically
unfavored (see Fig. \ref{fig:landscape}).
As $\alpha$ is increased above $\alpha_c$, the smooth-texture
minimum becomes metastable and the unbinding of a defect pair is
the most likely scenario.

\begin{figure}
\includegraphics[width=0.47\textwidth]{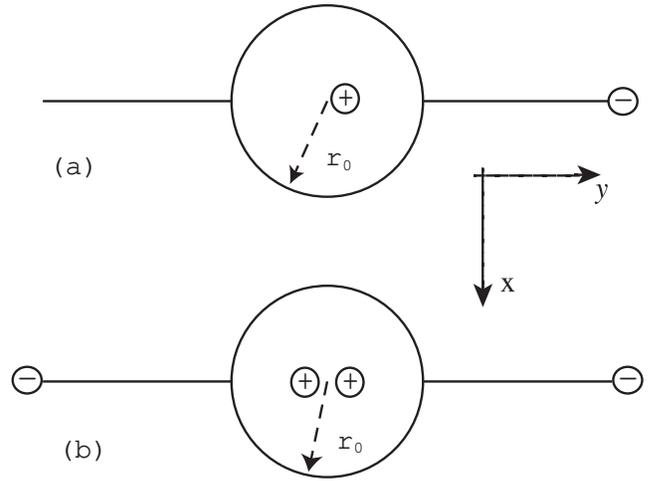}
\caption{\label{schematic} The equilibrium defects positions are
illustrated schematically in the case of one (a) and two dipoles
(b). We assume free boundary conditions at infinity, as in Fig.
\ref{defect free texture}, so that the effect of image charges can
be neglected.}
\end{figure}

\begin{figure}
\includegraphics[width=0.47\textwidth]{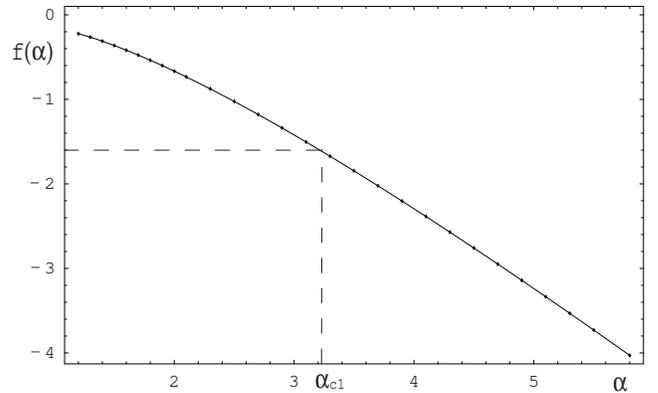}
\caption{Plot of $f(\alpha)$ versus the aspect ratio $\alpha$
obtained by minimizing over the single-dipole defect configuration
represented schematically in Fig. \ref{schematic}a. As discussed
in the text, the first unbinding transition occurs when $f(\alpha
_{c_{1}})$ is equal to $-\frac{q^2}{2 \pi}\ln\left(
\frac{r_{0}}{a'}\right)$. The value $\alpha_{c1}$ is indicated by
the dashed line for $\frac{r_{0}}{a'}=10^4$ and $q=\frac{2
\pi}{6}$ . Note that no minimum (corresponding to an unbound pair)
exists for $\alpha$ less than 1 (approximately), hence the curve
cannot be continued to the origin (see Fig \ref{fig:landscape}).}
\label{1dipole}
\end{figure}

It is useful to parameterize $\Delta F (\alpha)$ in terms of the
dimensionless radial coordinates $\bar{r}_i \equiv
\frac{r_{i}}{r_{0}}$. The geometric potential is defined in
Eq.(\ref{eq:pot-bis}) as a function of $\bar{r}_i$, $V(r)\equiv
\tilde{V}_{\alpha}(\frac{r}{r_{0}})$, where
\begin{eqnarray}
\tilde{V}_{\alpha}(x) = - \int_{x}^{\infty} \frac{dy}{y} \left(
\sqrt{1+\alpha^{2} y^{2} \exp(-y^{2})}-1 \right) \!\!\!\! \quad .
\label{norm2}
\end{eqnarray}
In order to write the defect-defect interaction in terms of the
dimensionless radial coordinate $\bar{r}_i$, we introduce a new
function $\tilde{\Re}(\bar{r}_i)$ defined by
\begin{eqnarray}
\tilde{\Re}(\bar{r}_i) &\equiv& \frac{r_{i}}{r_{0}}\exp[V(r_{i})]
=
\frac{r_{i}}{r_{0}}\exp \left[\tilde{V}_{\alpha}\left(\frac{r_{i}}{r_{0}}\right)\right] \nonumber\\
&=& \frac{\Re(r_{i})}{r_{0}}\!\!\!\! \quad , \label{norm}
\end{eqnarray}
where Eq.(\ref{solution2}) was used in the last step. We can now
transform $\Gamma_{a}(\bar{r}_i,\phi_{i},\bar{r}_j,\phi_{j})$ by
eliminating $\Re(r_{i})$ in favor of $\tilde{\Re}(\bar{r}_i)$.
Thus we have using Equations (\ref{green-cut}) and
(\ref{green2bis})
\begin{eqnarray}
-\sum_{j \neq i}^{N_{d}} q_{i} q_{j} \Gamma_{a}(r_i,\phi_{i},r_j,\phi_{j})&=& - \sum_{j \neq i}^{N_{d}} q_{i} q_{j} \Gamma(\bar{r}_i,\phi_{i},\bar{r}_j,\phi_{j}) \nonumber\\
&+& \frac{1}{2 \pi}\sum_{i=1}^{N_{d}}q_{i}^{2}\ln(\frac{r_{0}}{a})
\!\!\!\! \quad , \label{int-delta m}
\end{eqnarray}
where we have exploited charge neutrality and Eq.(\ref{green2}).
The free energy minimized with respect to the positions of the
defects, $min \!\!\!\!\! \quad \left[\frac{\Delta
F}{K_{A}}\right]$, can now be written, according to Eq.(\ref{delta
f}), as
\begin{eqnarray}
min \!\!\!\!\! \quad \left[\frac{\Delta F}{K_{A}}\right] =
f(\alpha)+\frac{1}{4 \pi} \sum_{i=1}^{N_{d}}q_{i}^{2}
\ln\left(\frac{r_{0}}{a'}\right) \!\!\!\! \quad , \label{delta
fm1}
\end{eqnarray}
where
\begin{eqnarray}
f(\alpha) &\equiv& min \!\!\!\! \quad [\frac{1}{2}\sum_{j \neq
i}^{N_{d}} q_{i} q_{j} \Gamma(\bar{r}_i,\phi_{i},\bar{r}_j,\phi_{j}) \nonumber\\
&+& \sum_{i=1}^{N_{d}}q_{i}\left(1-\frac{q_{i}}{4
\pi}\right)V(\bar{r}_i)] \!\!\!\! \quad . \label{delta m}
\end{eqnarray}
Note that in the second term in Eq.(\ref{delta fm1}) the core
energy of each defect has been absorbed in the modified core
radius $a'$, as defined in Eq.(\ref{a prime}). This generates an
energy cost for unbinding that can be overcome if $f(\alpha)$
assumes sufficiently large negative values. Hence it is sufficient
to study numerically how $\Delta F (\alpha)$ varies as a function
of a single parameter, eg. $\frac{r_{0}}{a'}$. As an illustration
of this approach, we study explicitly the unbinding of one and two
disclinations pairs leading to the ground states represented
schematically in Fig. \ref{schematic}.
The smooth ground state becomes unstable to the formation of one
defect dipole first. The critical aspect ratio above which this
scenario occurs can be determined with the aid of Fig.
\ref{1dipole}
where the function $f(\alpha)$ introduced in Eq.(\ref{delta m}) is
plotted as a function of the aspect ratio. From Eq.(\ref{delta
fm1}) we see that $\alpha_{c1}$ is determined by
\begin{equation}
f(\alpha _{c_{1}})= -\frac{q^2}{2 \pi}\ln\left(
\frac{r_{0}}{a'}\right) \!\!\!\! \quad . \label{int-delta m2}
\end{equation}
For $\frac{r_{0}}{a'}=10^{4}$ and $q=\frac{2 \pi}{6}$, we obtain a
critical aspect ratio $\alpha_{c_{1}} \approx 3.2$. As a
comparison, the approximate condition derived in Eq.(\ref{delta
f-ap}) gives $\alpha_{c_{1}}\approx 3$ when used in conjunction
with Fig. \ref{Vo}. The rougher estimate in Eq.(\ref{delta f-3})
leads (for $q=\frac{2\pi}{6}$) to $\alpha_{c_{1}}\approx 2.6$.
This discrepancy is easily understood considering that
Eq.(\ref{delta f-3}) was derived by means of a low $\alpha$
expansion.
\begin{figure}
\includegraphics[width=0.47\textwidth]{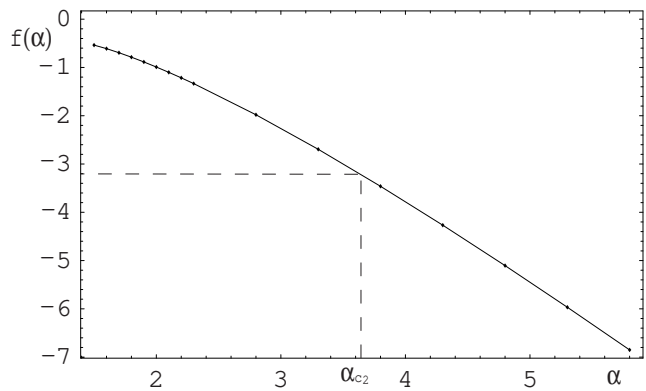}
\caption{\label{2dipole} Plot of $f(\alpha)$ versus the aspect
ratio $\alpha$ obtained by minimizing the two-dipole defect
configuration represented schematically in Fig. \ref{schematic}b .
The second unbinding transition occurs when $f(\alpha _{c_{2}})$
is equal to $-\frac{q^2}{\pi}\ln\left( \frac{r_{0}}{a'}\right)$.
The value $\alpha_{c_{2}}$ is indicated by the dashed line for
$\frac{r_{0}}{a'}=10^4$ and $q=\frac{2 \pi}{6}$ (compare with Fig.
\ref{1dipole}).}
\end{figure}

\begin{figure}
\includegraphics[width=0.47\textwidth]{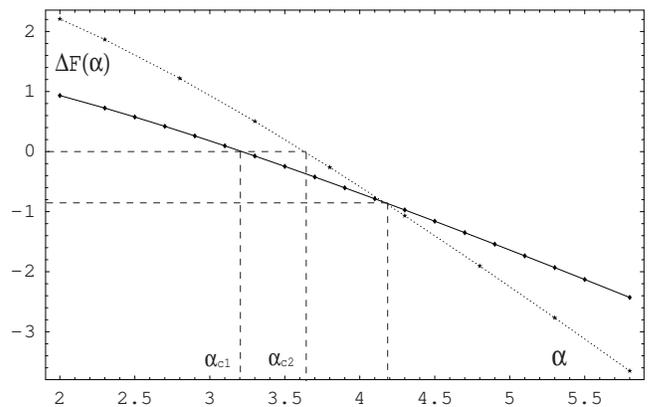}
\caption{\label{bothdipoles} Plot of $\frac{\Delta F
(\alpha)}{K_{A}}$ versus $\alpha$ corresponding to a single dipole
(continuous line) and two dipoles (dotted line) for
$\frac{r_{0}}{a'}=10^4$. The critical aspect ratios
$\alpha_{c_{1}}$ and $\alpha_{c_{2}}$ are indicated by dashed
lines. Note that the aspect ratio for which the two dipole
configuration becomes energetically favored occurs for
$\alpha>4.2$.}
\end{figure}

The critical aspect ratio $\alpha_{c_{1}}$ is too low for the
two-dipole defect configuration to become energetically favorable
with respect to the smooth ground state. Indeed, inspection of
Fig. \ref{2dipole} reveals that the critical aspect ratio
$\alpha_{c_{2}}$ for which the "two-dipole instability" sets in is
approximately equal to $3.6$ for the same choice of parameters
used in the single pair case. Note that, in the presence of two
dipoles, the energy cost arising from the second term in
Eq.(\ref{delta fm1}) is twice as large because there are four
defects rather than two. However, for $\alpha \gtrsim 4.2$
generating two dipoles becomes more energetically favored than a
single dipole (see Fig. \ref{bothdipoles}). The approach
illustrated here can be used to calculate a cascade of defect
unbinding instabilities at critical aspect ratios $\alpha_{c_{i}}$
involving higher number of dipoles and their equilibrium
configurations in the ground state. Note that the unbinding
eventually stops since the integrated Gaussian curvature in the
top cup cannot exceed $2 \pi$.  We expect the qualitative features
of our analysis to be independent of the exact shape of the bumpy
substrate although the specific values $\alpha_{c_{i}}$ depend on
the geometry and the choice of the microscopic parameters $a$ and
$E_c$. Finally, we emphasize that the curvature-induced unbinding
is similar to a first order transition and occurs for rather
pronounced deviations from flatness (ie. large $\alpha$).
\subsection {\label{single}Single vortex instability}
The unbinding of defect pairs may not be the most likely scenario
if the size of the system $R$ is sufficiently small. In this case,
the creation of a single vortex at the center of the bump may
become energetically favorable for lower aspect ratios than
required by the defect dipole instability. The equation for the
bond angle field $\theta_{s}({\bf u})$ for a single defect of
charge $q$ at the center of the bump is given by:
\begin{equation}
\theta_{s}(\phi) = \left(\frac{q}{2 \pi}-1\right)\phi \!\!\!\!
\quad , \label{sing-vortex}
\end{equation}
where the bond angle is measured with respect to the rotating
basis vectors corresponding to the polar coordinates discussed in
Section \ref{defe}. Upon substituting $\theta_{s}(\phi)$ in
Eq.(\ref{eq:patic-ener}) and subtracting the free energy ${\cal
F}_{0}$ corresponding to the defect-free texture we obtain:
\begin{eqnarray}
\frac{\Delta F (\alpha)}{K_{A}} &=& \frac{q^{2}}{4
\pi}\ln\left(\frac{\Re(R)}{a}\right) +
q\left(1-\frac{q}{4\pi}\right)V(0) \nonumber \\ &+& q^{2}
\frac{E_{c}}{K_{A}} \!\!\!\!\! \quad , \label{delta f-2}
\end{eqnarray}
where $E_{c}$ was added by hand. The same result is obtained by
using the more general formalism developed in Appendix D. Indeed,
by letting the position of an isolated defect tend to the center
of the bump in Eq.(\ref{eq:longD}) we obtain the energy of the
singular field in the case of free boundary conditions and the
result matches Eq.(\ref{delta f-2}). As discussed in Appendix
\ref{appD}, a defect located at $r_i$ is attracted to the boundary
at $R$ for free boundary conditions. One can think of this
interaction as resulting from an image defect of opposite sign
behind the edge of the sample at position $r'_{i}$ such that the
following relation holds in terms of the conformal radius
$\Re(r')$
\begin{eqnarray} \Re(r'_{i})=\frac{\Re(R)^{2}}{\Re(r_{i})}
\!\!\!\! \quad . \label{image2}
\end{eqnarray}
This result can be understood by analogy to the familiar
electrostatic problem of a charged line located a distance $r_{i}$
from the center of a cylindrical grounded conductor whose axis is
parallel to it \cite{Pano-book}. The analogy becomes precise if
one lets $r_{i}\rightarrow \Re(r_{i})$ as explained in Appendix
\ref{appD}.

If the geometric potential is not strong enough (as in the flat
space limit $\alpha=0$), the defect will migrate to the edge of
the sample and annihilate with its image leaving a smooth field.
On the other hand, when the aspect ratio is sufficiently large,
the defect can lower its energy by sitting at the center of the
bump. Comparison of Eq.(\ref{delta f-2}) with Eq.(\ref{delta
f-1-bis}) shows that, unless $R \gg r_{0}$, the energy of the
single vortex instability will be lower or at least comparable to
the unbinding of a defect dipole. In fact, the threshold
$\alpha_{s}$ that $\alpha$ needs to exceed to trigger the single
defect instability is easily obtained if the values of the
geometric potential at the origin are tabulated for different
aspect ratios, as illustrated in Fig. \ref{fig:pot}. The condition
for single vortex generation reads
\begin{eqnarray}
|V(0)|>\frac{q}{(q-4\pi)} \ln\left(\frac{R}{a'}\right) \!\!\!\!\!
\quad . \label{delta f-ap-s}
\end{eqnarray}
Using the same method adopted to derive Eq.(\ref{delta f-3}) we
obtain an estimate of how $\alpha_{s}$ depends on $\frac{R}{a'}$
(compare with Eq.(\ref{delta f-3})):
\begin{eqnarray}
\alpha_{s}^{2} \approx \frac{4q}{(4\pi-q)}
\ln\left(\frac{R}{a'}\right) \!\!\!\!\! \quad . \label{delta
f-3-bis}
\end{eqnarray}
The single vortex instability is reminiscent of vortex generation
in rotating superfluid helium with $\alpha$ playing the role of
the angular speed $\Omega$. For a volume of helium contained in a
cylindrical vessel of radius $R$ and rotating uniformly with
constant angular speed, the critical value $\Omega_{c_{1}}$ above
which defect generation occurs is given by \cite{Vine}
\begin{eqnarray}
\Omega_{c_{1}} \approx \frac{K}{2\pi R^{2}}
\ln\left(\frac{R}{a}\right) \!\!\!\!\! \quad , \label{helium}
\end{eqnarray}
where $K=\frac{2\pi\hbar}{m_{He}}$ is the magnitude of the quantum
of circulation and $a$ the core radius \footnote{A similar
mechanism applies to superconductors in a uniform magnetic
field.}. Note that $\Omega_{c_{1}}$ decreases as $R$ increases,
unlike $\alpha_{s}$ which diverges logarithmically. Thus, the
single defect instability studied here is a finite size effect. In
contrast, the disclination unbinding studied earlier in this
section does not depend on the system size because of charge
neutrality. Hence the thermodynamic limit can be safely taken,
provided the characteristic length over which the curvature varies
(ie. $r_{0}$) is not too large compared to $a$ (see Eq.(\ref{delta
f-3})).

In considering the case of small system size, it is important to
keep in mind two assumptions implicit in the present treatment.
The radius of curvature $\frac{r_{0}}{\alpha}$ must be much larger
than the core radius everywhere for the continuum approach to be
valid, that is $r_{0}\gg \alpha \!\!\!\!\! \quad a $.
Additionally, the Gaussian curvature must be vanishing small at
the edge of the system which requires $R$ to be larger than a few
$r_{0}$.

\subsection {Lattice of bumps, valleys and saddle points}

In some experimental realizations perhaps modelled on those of
Ref. \cite{sega03} the topography will be periodic. In this
Section we discuss qualitatively how the results described above
generalize to a 2-dimensional lattice of bumps with variable
aspect ratio for both square and triangular lattices. A more
quantitative approach to this problem would involve finding
conformal set of coordinates for periodic boundary conditions.
This is possible in principle but more involved since cylindrical
symmetry is now lost. Nonetheless, the intuition gained by
studying the single bump allows us to make some guesses for the
ground state. We first note that the geometric potential generated
by the lattice of bumps is not simply the superposition of results
for single bump potentials. This is caused by the non linear
relation between the surface height and the Gaussian curvature
acting as a source for the geometric potential. To explore this
point further, consider what happens when four bumps are placed at
the vertices of a square. At the center of the square a minimum of
the height function occurs corresponding to a $new$ region of
positive Gaussian curvature. This effect is particularly acute for
$r_0\leq L$, where $L$ is the bump spacing. In general
interference between bumps creates a dual lattice of valleys. A
similar breakdown of the superposition principle arises for
triangular lattices.
\begin{figure}
\includegraphics[width=0.45\textwidth]{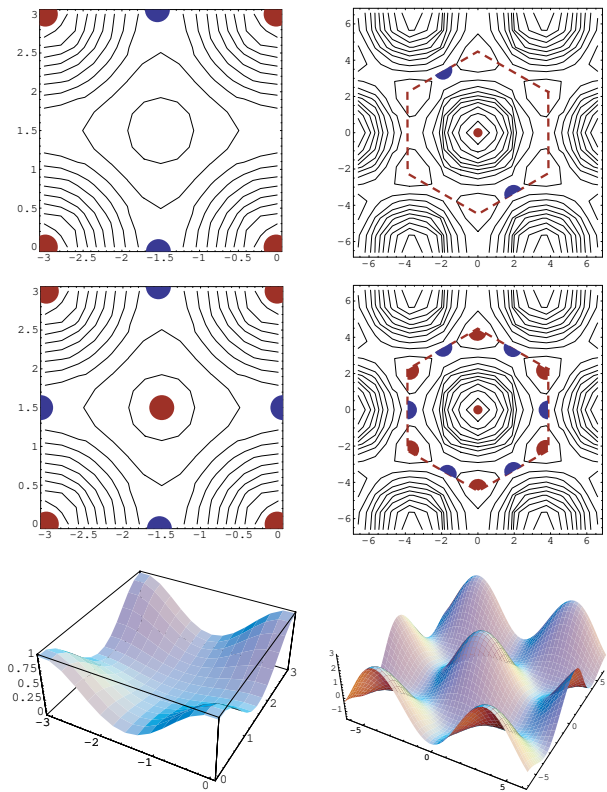}
\caption{\label{bumps array}(Top) Ground states for square (left)
and triangular (right) arrays of bumps. The first and second rows
correspond to moderate values of the aspect ratio $\alpha$
respectively. For simplicity, we assume that $r_0$, the bump width
is comparable to the lattice spacing. Positive defects (red dots)
"screen" regions of positive Gaussian curvature while negative
ones (blue dots) are located on the saddles of the "hilly"
landscape.}
\end{figure}

As the aspect ratio of hilly landscapes such as those shown in
Fig. \ref{bumps array} is increased, defects can be created to
screen the Gaussian curvature. Their positions can be guessed by
considering a unit cell of the lattice such that the integrated
Gaussian curvature vanishes. For a square lattice, we conjecture
that the first topography induced transition is associated with
the appearance of positive defects at the top of the bumps and
negative ones half way between them in the vertical or horizontal
direction (see Fig. \ref{bumps array}). This two-fold degeneracy
is compatible with the symmetry of the lattice and analogous to
the freedom in choosing the axis along which the first
disclination-dipole appears on the single bump. The negative
defects are shared between two adjacent cells while the positive
ones are shared among four cells thus ensuring overall charge
neutrality. As the value of $\alpha$ increases even more, one
might expect an additional positive defect appears in the valley
located at the center of each cell and two additional negative
defects shared with the adjacent cells are created between the
bumps at right angles to the direction discussed above (see Fig.
\ref{bumps array}).

For the triangular lattice, we conjecture that the first
transition corresponds to positive defects on top of the bumps and
negative ones between the bumps along one of the three axis of
symmetry of the unit cell. As the value of the aspect ratio is
increased, additional positive defects appear on the six minima of
the surface and negative ones are generated  along the remaining
two axes of symmetry of the unit cell (see Fig. \ref{bumps
array}). A simple count of the total defect charges enclosed in
the unit cell shows that this scenario also satisfies the
requirement of defect charge neutrality.
\section {\label{circ} Defect Deconfinement}
With potential experiments in mind \cite{Kram}, it is interesting
to consider the case of hexatic order on a bump encircled by a
circular wall of radius $R\gg r_0$ which aligns the hexatic bond
angles. As a simple model, imagine an array of hexagons which
locally achieve a common orientation tangential to the wall  (see
Fig. \ref{fig:appD1}b). The hexatic order parameter will thus
rotate by $2\pi$ upon making a circuit of the wall, insuring that
at least six defects of "charge" $\frac{2 \pi}{6}$ must be
included in the ground state for all values of the aspect ratio.
These boundary-condition induced defects will interact with the
Gaussian curvature of the bump and with the wall. The $N_{d}$
defects contribute large (constant) self energies of the form
$\sum_{i=1}^{N_{d}}K_{A} q_{i}^{2}\ln\left(\frac{R}{a}\right)$
that dominate the total energy for sufficiently large systems.
Since  $\sum_{i=1}^{N_{d}}q_{i}$ must be equal to $2 \pi$, the
energy is minimized when the defects split up into the smallest
possible charges.

The equilibrium defect configuration must minimize the free energy
taking into account the confining potential generated by the
Gaussian curvature and the interactions of the defects with the
boundary and among themselves. The repulsive force exercised by
the wall on a defect located at $\Re(r_{i})$ in the conformal
plane can be computed by placing an image defect of same charge
outside the wall at position $\frac{\Re(R)^2}{\Re(r_{i})}$. The
mathematics resembles the problem of finding the magnetic field of
a line current located at a given distance $r_i$ from the center
of a cylinder of high permeability material and whose radius $R$
is greater than $r_i$ \cite{Pano-book}. The analogy is complete
upon performing the change of coordinates $r \rightarrow \Re(r)$
and identifying the gradient of the bond angle
$\partial_{\alpha}\theta({\bf u})$ with the magnetic field. This
is explained in detail in Appendix \ref{appC} and \ref{appD} where
we introduce a conjugate function $\chi({\bf u})$ analogous to the
vector potential that simplifies the analysis of this problem.
Thus, each of the $N_d$ defects will also interact with an equal
number of image defects. This situation can be described
mathematically by deriving an appropriate Green's function
$\Gamma^{N}$ that includes the images, as discussed in Appendix
\ref{appD} (see Eq.(\ref{eq:wideeq})). The resulting free energy
$F^{N}$ reads:
\begin{eqnarray}
\frac{F^{N}}{K_{A}} &=& \frac{1}{2}\sum_{j \neq i}^{N_{d}} q_{i}
q_{j} \!\!\!\!\! \quad \Gamma^{N}(x_{i};x_{j}) + F_{0}
\nonumber\\&+& \sum_{i=1}^{N_{d}}q_{i}(1-\frac{q_{i}}{4
\pi})V(r_{i})+ \sum_{i=1}^{N_{d}} \frac{{q_{i}}^{2}}{4 \pi} \ln
\left[\frac{\Re(R)}{a}\right]\nonumber\\&-& \sum_{i=1}^{N_{d}}
\frac{{q_{i}}^{2}}{4 \pi} \ln \left[1-x_{i}^{2}\right] \!\!\!\!
\quad , \label{eq:longDbis}
\end{eqnarray}
where $F_{0}$ is defined in Eq.(\ref{free texture energy}) and the
Green's function $\Gamma^{N}(x_{i};x_{j})$ is given by
\begin{eqnarray}
\Gamma^{N}(x_{i};x_{j})=-\frac{1}{4 \pi} \ln
\left[x_{i}^{2}+x_{j}^{2}-2x_{i} x_{j} \cos \left( \phi_{i}-
\phi_{j}\right) \right]\nonumber\\ - \frac{1}{4 \pi} \ln
\left[x_{i}^{2} x_{j}^{2} + 1 - 2 x_{i} x_{j} \cos \left(\phi_{i}-
\phi_{j}\right) \right] \!\!\!\! \quad . \nonumber\\
\label{eq:green-norm2}
\end{eqnarray}
The last term accounts for the interaction with the image defects
and the superscript $N$ indicates Neumann boundary conditions on
an appropriate potential function. Here, we use scaled coordinates
in the conformal plane $x_{i} \equiv \frac{\Re(r_{i})}{\Re(R)}$.
The interaction of the defects with the curvature is not affected
by the presence of the distant wall.

\begin{figure}
\includegraphics[width=0.45\textwidth]{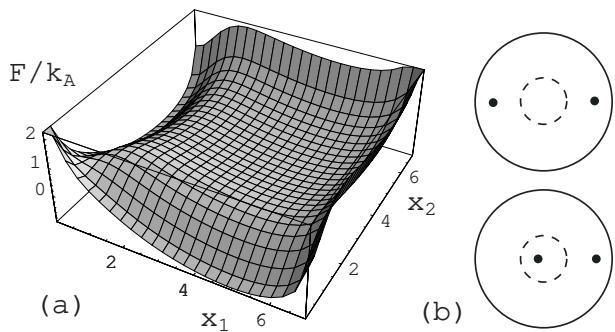}
\caption{\label{fig:nematic} (a) The free energy for a nematic
(double headed vector field) living on a Gaussian bump surrounded
by an aligning circular wall is plotted for $\alpha=2$ as a
function of the scaled radial coordinates $x_{1}$ and $x_{2}$ of
the two disclinations. The radial coordinates have been scaled by
$r_{0}$ and the size of the system is $R=7 r_{0}$. Note that the
energy plot is symmetric with respect to the line $x_{1}=x_{2}$.
(b) Schematic illustration of the positions of the two
disclinations (black dots) corresponding to the deep energy minima
at positions $x_{1}=0.04$ and $x_{2}=4.9$ (or viceversa) and to a
shallow minimum at $x_{1}=x_{2}=4.7$. The two defects are on
opposite sides of the bump. The continuous line corresponds to the
circular boundary while the dashed one to the circle of zero
Gaussian curvature and radius $r_{0}$ (drawing not to scale).}
\end{figure}

To provide an illustration of the combined effect of curvature and
boundary conditions on tangential vector order, we first consider
the simpler case of a nematic order parameter with periodicity
equal to $\pi$. This simplified model neglects differences in the
elastic constants for bend and splay and does not incorporate any
effect due to the uniaxial coupling of the nematogens to the
curvature. In this case, minimization of the logarithmically
diverging part of the free energy (fourth term in Eq.
\ref{eq:longDbis}) suggests that there will be only two
disclinations of charge $q=\pi$ displaced along a radial direction
(see Fig. \ref{fig:nematic}b). By applying Eq.(\ref{eq:longDbis}),
we can parameterize the energy of the system in terms of the
scaled radial coordinates, $x_{1}$ and $x_{2}$ of the two
disclinations. The resulting energy landscape is plotted in Fig.
\ref{fig:nematic} (for $\alpha = 2$ and $R=7 r_0$) and clearly
reveals two minimal-energy configurations. The first minimum
corresponds to one disclination confined at the top of the bump
(slightly shifted from the center) and the other at a radial
distance approximately 70\% of $R$ (see Fig. \ref{fig:nematic}b
bottom panel). The second minimum corresponds to a fully
deconfined state with both disclinations placed symmetrically at
approximately 67\% of $R$ (see Fig. \ref{fig:nematic}b top panel).
As the aspect ratio is raised even further, the saddle in the
energy landscape of Fig. \ref{fig:nematic}a becomes a minimum
corresponding to a configuration in which both disclinations are
confined in the cup of positive Gaussian curvature by the
geometric potential.

\begin{figure}
\includegraphics[width=0.45\textwidth]{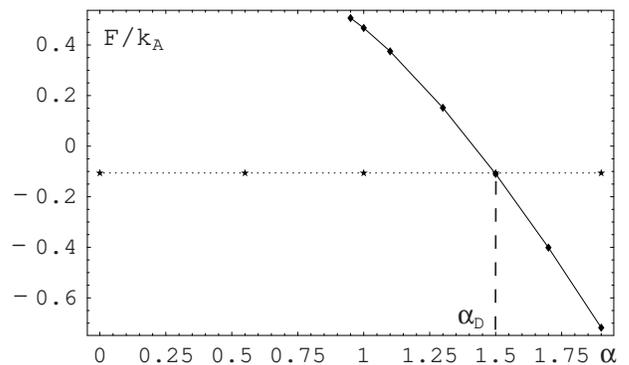}
\caption{\label{fig:deconftrans} Plot of the free energy of a
nematic (double headed vector field) on a Gaussian bump encircled
by an aligning wall as a function of $\alpha$. The dotted line
represents the energy of the fully deconfined configuration in
Fig. \ref{fig:nematic}b top panel while the continuous line
corresponds to the defect pattern illustrated in the bottom panel
of Fig. \ref{fig:nematic}b. The energy of the fully deconfined
configuration is approximately independent of $\alpha$ because the
two disclinations are far away from the bump.}
\end{figure}

As illustrated in Fig. \ref{fig:deconftrans}, there is a critical
value of the aspect ratio, $\alpha_{D} \simeq 1.5$, above which it
is energetically favorable for the system to have one disclination
confined at the top of the bump. For $\alpha < \alpha_{D}$ the
fully deconfined configuration becomes energetically favorable,
but the two minima can still coexist. As $\alpha$ is decreased
even further, the repulsion between the two disclinations
overcomes the confining force of the geometric potential and makes
the second minimum in Fig. \ref{fig:deconftrans}a (corresponding
to the partially confined configuration) disappear altogether.
This "spinodal point" occurs for $\alpha \approx 0.9$ on the
Gaussian bump. The specific values of the critical aspect ratios
are geometry dependent, but the generic mechanism of deconfinement
depends only on a large separation of the length scales $r_{0}$
and $R$ that control the interaction with the curvature and the
boundary respectively.

\begin{figure}
\includegraphics[width=0.47\textwidth]{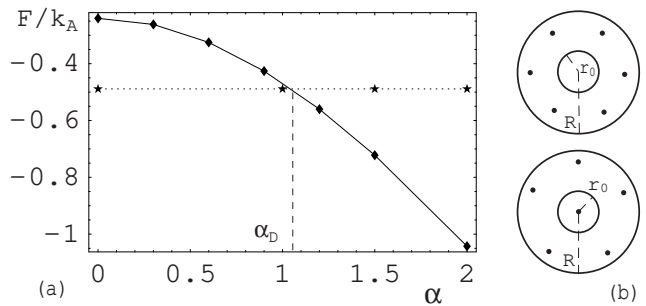}
\caption{\label{fig:hexpent} Plot of the free energy of an hexatic
phase (draped on the Gaussian bump encircled by a wall) as a
function of $\alpha$. The dotted line represents the energy of the
hexagonal configuration illustrated in the top panel on the left
while the continuous line corresponds to the pentagonal
arrangement in the bottom panel. The outer defect rings in both
configurations are approximately 90\% of $R$. The critical aspect
ratio $\alpha_{D}$ corresponding to the deconfinement transition
discussed in the text is indicated by the dashed line.}
\end{figure}

The analysis for the hexatic case is complicated by the fact that
more defect configurations are possible when six defects are
present. We start by noting that even in flat space ($\alpha = 0$)
there are two natural low energy defect configurations with high
symmetry: the ground state corresponding to the six defects
sitting at the vertices of an hexagon and a higher energy state
given by a pentagonal distribution of defects with the sixth
defect sitting at the center of the circular sample (see Fig.
\ref{fig:hexpent}b). As the aspect ratio is raised, the pentagonal
arrangement becomes energetically favored since it pays to have a
defect confined in the (geometric) potential well at the origin
(see Fig. \ref{fig:hexpent}a). To study the transition, it is
useful to derive expressions for the energy of the two defect
configurations as a function of the radius of the outer defect
ring $r$. Every defect (except the one at the origin, possibly)
has the same scaled coordinate $x_{i}=x$ and the angles between
two defects are integer multiples of $\frac{2 \pi}{n}$ where $n$
is the number of defects in the outer ring ($n=5$ for the pentagon
and $n=6$ for the hexagon). In this case, the sums involved in the
first (interaction) term in Eq.(\ref{eq:longDbis}) can be
efficiently evaluated using the following identity:
\begin{eqnarray}
\frac{1}{2} \sum_{i}^{n-1} \ln \left[p^2 + 1 - 2p \cos(\frac{2 \pi
i}{n}) \right] \nonumber\\ = \ln \left( 1-p^n \right)-
\ln\left(1-p \right) \!\!\!\! \quad . \label{eq:ident2}
\end{eqnarray}
Upon using Eq.(\ref{eq:ident2}) with $p=1$ and $p=x^{2}$ to
evaluate the sums arising from the first and the second term of
the Green's function in Eq.(\ref{eq:green-norm2}) respectively, we
obtain the free energy $F_{H}$ for the hexagonal configuration:
\begin{eqnarray}
\frac{F_{H}(\alpha)}{K_{A}} &=& - \frac{ \pi}{6} \!\!\!\!\! \quad
\ln\left[\left(\frac{\Re(r)}{\Re(R)}\right)^{5}-\left(\frac{\Re(r)}{\Re(R)}\right)^{17}\right]-
\frac{ \pi}{6} \!\!\!\!\! \quad \ln6 \nonumber\\ &+& \frac{11
\pi}{6} \!\!\!\! \quad V(r) + \frac{ \pi}{6} \!\!\!\!\! \quad
\ln\left[\frac{\Re(R)}{a}\right] \!\!\!\! \quad . \label{eq:hex}
\end{eqnarray}
The free energy for the pentagonal configuration $F_{P}$ is
readily obtained after similar manipulations
\begin{eqnarray}
\frac{F_{P}(\alpha)}{K_{A}} &=& - \frac{5 \pi}{36} \!\!\!\!\!
\quad
\ln\left[\left(\frac{\Re(r)}{\Re(R)}\right)^{6}-\left(\frac{\Re(r)}{\Re(R)}\right)^{16}\right]-
\frac{ \pi}{2} \!\!\!\!\! \quad \ln2 \nonumber\\ &+& \frac{11
\pi}{36} \!\!\!\! \quad V(0) + \frac{55 \pi}{36} \!\!\!\! \quad
V(r) + \frac{ \pi}{6} \!\!\!\!\! \quad
\ln\left[\frac{\Re(R)}{a}\right] \!\!\!\! \quad . \label{eq:pent}
\end{eqnarray}
Note that these manipulations are very similar to the ones
necessary to describe superfluid helium in a cylinder of radius
$R$ \cite{hess67}. In fact, the superfluid problem is analogous to
the case of hexatic order with $free$ boundary conditions on a
circular boundary of radius $R$ (see Appendix \ref{appD}). The
rather unusual form of the argument of the logarithm in Equations
(\ref{eq:hex}) and (\ref{eq:pent}) arises from the sum over the
image defects whose positions depend non-linearly on the position
of the defects themselves.

Minimization of Equations (\ref{eq:hex}) and (\ref{eq:pent}) with
respect to $r$ fixes the distance of the outer defects. The
resulting minimal energies $F_{P}$ and $F_{H}$ are plotted as a
function of $\alpha$ in Fig. \ref{fig:hexpent}a.
The critical value $\alpha_{D}$, for which $F_{P}< F_{H}$ can be
easily estimated by realizing that $F_{H}$ is approximately
independent of $\alpha$ because the disclinations are far from the
bump. On the other end, $F_{P}$ decreases with $\alpha$ because
the confined disclination is trapped in a potential well whose
depth is approximately given by $-\frac{11}{144} \!\!\!\! \quad
\pi \alpha ^{2}$ (see the second term of Eq.(\ref{eq:pent}) and
the low $\alpha$ expansion for $V(0)$ derived in
Eq.(\ref{potential-approx})). The critical aspect ratio,
$\alpha_{D}$, for which the deconfinement transition occurs can be
estimated by setting the depth of this potential well equal to the
energy difference between the hexagon and pentagon configurations
in flat space. The latter can be read off from the energy diagram
in Fig. \ref{fig:hexpent}a and the result is approximately $0.3
K_{A}$ which leads $\alpha_{D}\sim 1.1$ in agreement with the
value indicated in Fig. \ref{fig:hexpent}a.

As the aspect ratio is raised even further, less symmetric defect
configurations become energetically favored corresponding to a
larger number of disclinations confined in the cup of positive
Gaussian curvature. For example, when two disclinations are
confined within $r=r_{0}$, the outer defect ring is given by four
defects approximately located at the vertices of a square. We note
that these defect configurations cease to exist at low aspect
ratios because they require the geometric potential to overcome
the strong repulsive interaction between the confined defects. As
discussed earlier for Fig.(\ref{fig:deconftrans}), the actual
values of the aspect ratios involved depend on the specific
geometry of the substrate. However the basic mechanism behind the
deconfinement transition is more general.

Note that, as $\alpha$ increases, the geometric mechanism of
defect dipole unbinding discussed in the last section may also set
in. Because of the presence of one or more positive defects at the
top of the bump, the critical aspect ratio necessary to unbind one
dipole will be larger than what was calculated before. If dipole
unbinding does occur, the new defects will "decorate" the existing
patterns by adding new positively charged disclinations in the
region of positive Gaussian curvature and expelling the negative
ones in the external region of the bump ($r>r_0$) where the
Gaussian curvature is also negative (see Fig.\ref{bump}).

\section{\label{conc}Conclusion}

We have discussed how the varying curvature of a surface such as a
"Gaussian bump" can trigger the generation of single defects or
the unbinding of dipoles, even if no topological constraints or
entropic arguments require their presence. This mechanism is
independent of temperature if the system is kept well below its
Kosterlitz Thouless transition temperature. It would be
interesting to revisit Kosterlitz-Thouless defect unbinding
transitions on surfaces of varying Gaussian curvature in the
presence of a quenched topography \cite{sach84} in the light of
the present work. One might also explore the $dynamics$ of the
delocalization transition that occurs when a bump is confined by a
circular edge and the aspect ratio is lowered until the defects,
initially confined on top of the bump by the geometric potential,
are forced to "slide" towards the boundary. Quantitative studies
of periodic arrangements of bumps would be interesting and could
be inspired by fruitful analogies with methods and ideas from
solid state physics.

We also hope to extend this work by considering crystalline order
on bumpy topographies and taking explicitly into account the
screening of clouds of dislocations and possible generation of
grain boundaries \cite{Bowi98}. Such an analysis would facilitate
comparison with experiments performed with a single grain of block
copolymer spherical domains \footnote{The radius of
block-copolymer spherical cores is of the order of a few
nanometers and their spacing tens of nanometers. These values can
be tuned by suitably choosing the block-copolymers and varying
their volume fraction.} on a suitably patterned substrate
\cite{sega03,Kram}.

\begin{acknowledgments}
We wish to acknowledge helpful conversations with M. Bowick, B. I.
Halperin, A. Hexemer, E. Kramer, S. Minwalla, K. Papadodimas, A.
Travesset and A. Turner. This work was supported by the National
Science Foundation, primarily through the Harvard Materials
Research Science and Engineering Laboratory via Grant No.
DMR-0213805 and through Grant No. DMR-0231631.
\end{acknowledgments}

\appendix

\section{\label{appA}Green's function and Isothermal Coordinates}

\begin{figure}
\includegraphics[width=0.47\textwidth]{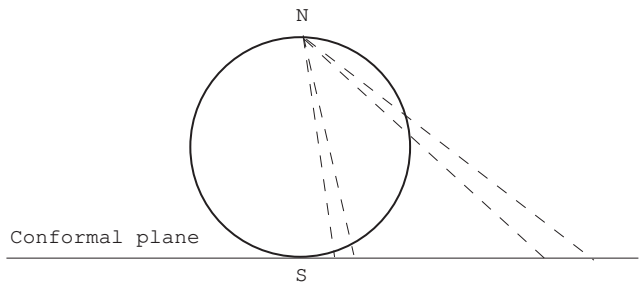}
\caption{\label{fig:stereo}Graphic construction of the
stereographic projection. Regions close to the north pole have
larger images in the conformal plane than regions of equal areas
close to the south pole. The stereographic projection preserves
the topology of the surface provided all points at infinity are
identified with the north pole.}
\end{figure}

The analysis of ordered phases on curved substrates can be
simplified by rewriting the original metric of the surface
$g_{ab}({\bold u})$ in terms of a convenient set of coordinates
${\bold x({\bold u})}=(x({\bold u}),y({\bold u}))$ such that the
new metric $\tilde{g}_{ab}({\bold x})$ reads
\begin{equation}
\tilde{g}_{ab}({\bold x})=e^{\rho(x,y)} \delta_{ab} \!\!\!\! \quad
. \label{metricappendix0}
\end{equation}
The metric $\tilde{g}_{ab}$ differs from the flat space one
$\delta_{ab}$ only by a conformal factor $e^{\rho(x,y)}$ that
embodies information on the curvature of the surface
\cite{Dubrovinbook}. These {\it isothermal coordinates} can be
used to map arbitrary corrugated surfaces onto the plane
\cite{Yp-book}. The mapping is conformal so angles are left
unchanged but areas are stretched according to the
position-dependent conformal factor $e^{\rho(x,y)}$. A familiar
example is provided by the stereographic projection that maps a
sphere onto the conformal plane as illustrated in Fig.
\ref{fig:stereo}.
The Green's function assumes a very simple form after a conformal
transformation, because the Laplace operator reduces to the
familiar flat space result when expressed in terms of isothermal
coordinates. In what follows, we demonstrate that this
transformation provides the basis for an efficient strategy to
determine the Green's function on a bumpy substrate.  We start by
deriving the radial change of coordinates $\Re(r)$ that transforms
the original metric of the Gaussian bump, i.e.,
\begin{equation}
ds^{2}=\left(1+ \frac{{\alpha}^2 r^2}{r_{o}^2}
e^{-\frac{r^2}{r_{o}^2}}\right) dr^{2} + r^2 d \phi^{2} \!\!\!\!
\quad , \label{metricappendix1}
\end{equation}
into the locally flat metric (in polar coordinates),
\begin{equation}
ds^{2}=e^{\rho(r)} (d \Re^{2} + \Re^{2} d \phi^{2})\!\!\!\! \quad
, \label{metricappendix2}
\end{equation}
where $\rho(r)$ and $\Re(r)$ are independent of the azimuthal
coordinate $\phi$ because of cylindrical symmetry. This metric is
equivalent to $\tilde{g}_{ab}({\bold x})$ upon switching from
cartesian (x,y) to polar coordinates $(\Re(r),\phi)$. To simplify
the notation we introduce the $\alpha$-dependent function $l(r)$
defined by
\begin{equation}
l(r) \equiv 1+ \frac{{\alpha}^2 r^2}{r_{0}^2} \exp
\left(-\frac{r^2}{r_{0}^2}\right) \!\!\!\! \quad , \label{l-bis}
\end{equation}
and plotted in Fig. \ref{fig:l(r)} for different choices of
$\alpha$.

The equivalence of the metrics in Equations
(\ref{metricappendix1}) and (\ref{metricappendix2}) requires that
$\Re(r)$ satisfies the differential equation
\begin{equation}
\frac{d\Re}{\Re}=\frac{\sqrt{l(r)}}{r}dr \!\!\!\! \quad .
\label{step}
\end{equation}
The conformal factor is thus given by
\begin{equation}
e^{\rho(r)}=(\frac{r}{\Re})^2 \!\!\!\! \quad . \label{sol-bis}
\end{equation}
The solution of Eq.(\ref{step}) is
\begin{equation}
\Re(r)= A \!\!\!\!\! \quad r \!\!\!\!\! \quad e^{-\int_{r}^{c}
\frac{dr'}{r'}(\sqrt{l(r')} \!\!\!\! \quad -1)} \!\!\!\! \quad ,
\label{solution}
\end{equation}
where it is convenient to set the arbitrary constants $A$ and $c$
to unity and infinity respectively. This non-linear stretch of the
radial coordinate leaves the origin and the point at infinity
invariant and can be concisely written as
\begin{equation} \Re(r)= r \!\!\!\!\!
\quad e^{V(r)} \!\!\!\! \quad , \label{solution2}
\end{equation}
where the function $V(r)$ defined by
\begin{equation}
V(r) \equiv - \int_{r}^{\infty} \! dr' \frac{\sqrt{l(r')}-1}{r'}
\!\!\!\! \quad , \label{eq:pot-bis}
\end{equation}
plays an important role in our formalism and its interpretation as
a sort of geometric potential is explored in detail in Appendix
\ref{appB}.

The Poisson equation for the Green's function $\Gamma({\bf u},{\bf
u'})$ on a surface with metric tensor $g_{\alpha \beta}$ and point
source $\delta({\bf u},{\bf u'})$ reads \cite{Davidreview}:
\begin{equation}
D^{\alpha}D_{\alpha}\Gamma({\bf u},{\bf u'}) = -\frac{\delta({\bf
u},{\bf u'})}{\sqrt{g}} \!\!\!\! \quad , \label{poisson}
\end{equation}
where the covariant Laplacian is given for general coordinates by:
\begin{equation}
D^{\alpha}D_{\alpha} \equiv (1/\sqrt{g})\partial_{\alpha}[\sqrt{g}
g^{\alpha\beta}\partial_{\beta}]\!\!\!\! \quad . \label{laplacian}
\end{equation}
The conformal change of coordinates transforms $\sqrt{ g(r,\phi)}$
into $e^{\rho(r)} \sqrt {g(\Re,\phi)}$ and
$g^{\alpha\beta}(r,\phi)$ into $e^{-\rho(r)}
g^{\alpha\beta}(\Re,\phi)$. The factors of $e^{\rho(r)}$ inside
the square brackets in Eq.(\ref{laplacian}) then cancel and we are
left with the flat space Laplacian in the polar coordinates
$(\Re(r),\phi)$. We conclude that $\Gamma({\bf u},{\bf u'})$ is
simply the Green's function of an undeformed plane expressed in
terms of the polar coordinates ($\Re(r),\phi)$:
\begin{equation} \Gamma({\bf u},{\bf u'})=-\frac{1}{4 \pi} \ln
[\Re(r)^{2}+\Re(r')^{2}-2\Re(r)\Re(r')\cos (\phi-\phi')] \!\!\!\!
\quad , \label{green2bis}
\end{equation}
where an arbitrary additive constant $C$ (which can be used to
satisfy boundary conditions at infinity) has been dropped. We note
that $\Gamma({\bf u},{\bf u'})$ differs from the flat space
Green's function by a non linear stretch of the radial coordinate.
In Appendix \ref{appC}, we will use $\Gamma({\bf u},{\bf u'})$ to
solve Poisson's equation on an infinite bumpy domain and calculate
the energy stored in the field. As in flat space, the Green's
function $\Gamma({\bf u},{\bf u'})$ will be suitably modified in a
finite system in a way that depends on the boundary conditions
chosen at the edge of the sample (see Appendix \ref{appD}).

We conclude this appendix by evaluating $\Gamma({\bf u},{\bf u'})$
when the two points ${\bf u}$ and ${\bf u'}$ are assumed to be
separated by a fixed distance $a$ small enough so that the surface
can be approximated by the local tangent plane to the Gaussian
bump. This short distance behavior will be useful when evaluating
the effect of a $constant$ core radius on the energetics of a
disclination at an arbitrary position. The fixed microscopic
length $a$ on the bump is stretched when projected in the
conformal plane (see, e.g., Fig. \ref{fig:stereo}) and assumes the
position dependent value $\lambda(x,y)$ given by
\begin{equation}
\lambda(x,y) = a e^{-\frac{\rho(x,y)}{2}} \!\!\!\! \quad .
\label{metappend}
\end{equation}
For a Gaussian bump cylindrical symmetry requires that $\lambda
(r,\phi)$ is dependent only on $r$ and can be explicitly written
upon using Equations (\ref{sol-bis}) and (\ref{solution2}) as
\begin{equation}
\lambda(r,\phi)= a \frac{\Re(r)}{r} = a e^{V(r)} . \label{lambda}
\end{equation}
We can now evaluate $\Gamma({\bf u},{\bf u'})$ in the limit ${\bf
u'} \rightarrow {\bf u + a}$, where this concise notation means
that the two points on the surface with coordinates ${\bf u}$ and
${\bf u'}$ are separated by an infinitesimal distance $a$ measured
on the bump. It does not matter in what direction the two points
approach each other as long as $a$ is small compared to the local
radii of curvature. Upon using Equations (\ref{green2bis}) and
(\ref{lambda}), we obtain
\begin{equation} \Gamma({\bf u},{\bf u}+{\bf a})=-\frac{1}{4 \pi} \ln
(a^{2}) - \frac{V(r)}{2 \pi} \!\!\!\! \quad . \label{green3}
\end{equation}
We note that $\Gamma({\bf u},{\bf u}+{\bf a})$ for fixed $a$ is
not a constant like in flat space but varies with position as the
function $V(r)$, reflecting the lack of translational invariance
on an inhomogeneous surface, where properties such as the Gaussian
curvature also vary with position.

\section {\label{appB}Geometric Potential}

In this appendix we present two equivalent ways of determining the
explicit form of the geometrical potential $V({\bf u})$ valid for
azimuthally symmetric surfaces like the Gaussian bump. The
starting point is the general definition introduced in Section
\ref{ener}:
\begin{equation}  V({\bf u}) \equiv -\int\! dA' \!\!\!\quad  G({\bf u}') \!\!\!\quad
\Gamma({\bf u},{\bf u}') \!\!\!\! \quad , \label{geometric
potential}
\end{equation}
with the Green's function $\Gamma$ as defined in
Eq.(\ref{poisson}). In the electrostatic analogy, $V({\bf u})$ is
thus the potential induced by a continuous distribution of
"charge" represented by the Gaussian curvature (with sign
reversed). We shall derive an analogue of Gauss law for corrugated
surfaces where the curvature of the surface (with sign reversed)
plays the role of a continuous density of electrostatic charge.

The first derivation makes use of the fact that $V({\bf u})$ is a
scalar under conformal transformations. This symmetry can be
checked explicitly by applying the same reasoning adopted for the
equation satisfied by the Green's function in Appendix \ref{appA}
to Eq.(\ref{geometric potential}). In fact, upon operating on both
sides of Eq.(\ref{geometric potential}) with the covariant
Laplacian and using Eq.(\ref{poisson}), the defining equation for
the geometric potential can be cast into the differential form:
\begin{equation}  D^{\alpha}D_{\alpha}V({\bf u}) = G({\bf u})\!\!\!\! \quad .
\label{eq-potent}
\end{equation}
The Gaussian curvature in Eq.(\ref{eq-potent}) can be written in
conformal coordinates \cite{Dubrovinbook} as
\begin{equation}
G(x,y)=-e^{-\rho(x,y)}(\partial^{2}_{x}+\partial^{2}_{y})\frac{\rho(x,y)}{2}
\!\!\!\! \quad , \label{metricappendix3}
\end{equation}
where $\rho(x,y)$ is the conformal factor introduced in Appendix
\ref{appA}. Similarly the left hand side of Eq.(\ref{eq-potent})
can be expressed in conformal coordinates \cite{Dubrovinbook} as
\begin{equation}
D^{\alpha}D_{\alpha}V({\bf
x})=+e^{-\rho(x,y)}(\partial^{2}_{x}+\partial^{2}_{y})V(x,y)
\!\!\!\! \quad . \label{metricappendix3-bis}
\end{equation}
Upon substituting Equations (\ref{metricappendix3}) and
(\ref{metricappendix3-bis}) in Eq.(\ref{eq-potent}), we conclude
immediately that the geometric potential in conformal coordinates
$V({\bf x})$ is given by
\begin{equation}
V(x,y)=-\frac{\rho(x,y)}{2} \!\!\!\! \quad .
\label{metricappendix4}
\end{equation}
Upon using Equations (\ref{sol-bis}), (\ref{solution2}) and
(\ref{eq:pot-bis}) to substitute in Eq.(\ref{metricappendix4}),
one obtains the explicit form of the geometric potential for the
bump parameterized by the coordinates $(r,\phi)$:
\begin{equation}
V(r)=-\int_{r}^{\infty} \! dr' \frac{\sqrt{l(r')}-1}{r'} \!\!\!\!
\quad , \label{potential4}
\end{equation}
where the $\alpha$-dependent function $l(r)$ was defined in
Eq.(\ref{l-bis}) and plotted in Fig. \ref{fig:l(r)}. The result of
the integration in Eq.(\ref{potential4}) is independent of $\phi$
because of azimuthal symmetry. The upper limit of integration is
chosen consistently with Eq.(\ref{eq-potent}) as usually done in
electrostatics.

A second derivation of this result is obtained by making explicit
use of the azimuthal symmetry of the bump and deriving a covariant
form of Gauss' law which allows an intuitive understanding of the
interaction between defects and curvature. This curved space
version of Gauss' law illuminates the electrostatic analogy used
throughout the text. The gradient of the geometric potential
defines an "electric field" $E_{\alpha}$ given by
\begin{equation}
E_{\alpha} \equiv - D_{\alpha} V({\bf u})= \int\! dA' \!\!\!\quad
G({\bf u}') \!\!\!\quad D_{\alpha} \Gamma({\bf u},{\bf u}')
\!\!\!\! \quad , \label{E-bis}
\end{equation}
where we have used Eq.(\ref{geometric potential}). One might
expect that the flux of the vector $E^{\alpha}$ through a closed
loop is proportional to the enclosed Gaussian curvature in analogy
with Gauss' law that relates the flux of the electric field to the
electrostatic charge enclosed. To prove this assertion, we invoke
the generalized Stokes' formula \cite{Davidreview} that relates
the surface integral over $A$ of the gradient of a field to its
flux through the contour loop $C$:
\begin{equation}   \int_{A}\! dA \!\!\!\quad
D_{\alpha} E^{\alpha} = - \oint_{C}\! d u^{\alpha} \gamma_{\alpha}
\!\!\!\!\!\! \quad ^{\beta} E_{\beta} \!\!\!\! \quad .
\label{Coulomb model2}
\end{equation}
The covariant antisymmetric tensor $\gamma_{\alpha \beta}$ is
given by
\begin{equation} \gamma_{\alpha \beta}=\sqrt{g} \epsilon_{\alpha \beta} \!\!\!\! \quad ,
\label{gamma}
\end{equation}
where $\epsilon_{\alpha \beta}$ is the anti-symmetric tensor with
$\epsilon_{r \phi}=-\epsilon_{\phi r}=1$. Similarly
$\gamma^{\alpha \beta}$ equals $\frac{\epsilon_{\alpha
\beta}}{\sqrt{g}}$ and the following identity holds \cite{Park96}
\begin{equation} \gamma^{\alpha \sigma}\gamma_{\sigma \beta}=-\delta^{\alpha}_{\beta} \!\!\!\!
\quad . \label{gamma1}
\end{equation}
The tensor $\gamma_{\alpha} \!\!\!\!\!\! \quad
^{\beta}=\gamma_{\alpha \sigma}g^{\sigma \beta}$ performs
anti-clockwise rotations of $\frac{\pi}{2}$ when acting on an
arbitrary tangent vector $V_{\beta}$, as can be checked by
evaluating $V^{\alpha}\gamma_{\alpha} \!\!\!\!\!\! \quad
^{\beta}V_{\beta}=\gamma_{\alpha \beta}V^{\alpha}V^{\beta}=0$,
where we have used the antisymmetry of $\gamma_{\alpha \beta}$.
Thus, the vector $d u^{\alpha} \gamma_{\alpha} \!\!\!\!\!\! \quad
^{\beta}$ in Eq.(\ref{Coulomb model2}) represents an infinitesimal
contour length times the inward unit vector perpendicular to it.
The dot product with the field $E_{\beta}$ then generates the
flux. To calculate the flux piercing a circular circuit centered
on a Gaussian bump, we will need to explicitly evaluate
$\gamma_{\phi} \!\!\!\!\!\! \quad ^{r}$,
\begin{equation}
\gamma_{\phi} \!\!\!\!\!\! \quad ^{r} = - \!\!\!\!\! \quad
\frac{r}{\sqrt{l(r)}} \!\!\!\!\! \quad . \label{gamma-exp}
\end{equation}

Upon using Eq.(\ref{E-bis}) we can rewrite the left hand side of
Eq.(\ref{Coulomb model2}) as
\begin{equation}   \int\! dA \!\!\!\quad
D_{\alpha} E^{\alpha} = \int\! dA \!\!\!\ \int\! dA' \!\!\!\quad
G({\bf u}') \!\!\!\quad D_{\beta}D^{\beta} \Gamma({\bf u},{\bf
u}') \!\!\!\! \quad . \label{step3}
\end{equation}
If we now recall the defining Equation (\ref{poisson}) of the
Green's function $\Gamma({\bf u},{\bf u}')$ and keep in mind that
the Laplacian in Eq.(\ref{step3}) operates on the variables
labelled by ${\bf u}'$ (not ${\bf u}$), we obtain using
Eq.(\ref{Coulomb model2}) a general result for the flux piercing a
closed loop on the surface, namely
\begin{equation}
\oint_{C}\! d u^{\alpha} \gamma_{\alpha} \!\!\!\!\!\! \quad
^{\beta} E_{\beta} = \int_{A} \! d^{2}u \sqrt{g} \!\!\!\quad
G({\bf u}) \!\!\!\! \quad . \label{Gauss}
\end{equation}
We can explicitely evaluate the right hand side of
Eq.(\ref{Gauss}) with the aid of Eq.(\ref{intgr Gauss curv}). By
appealing to the cylindrical symmetry, in the special case of the
Gaussian bump one can apply this covariant form of Gauss' theorem
to find the radial field $E^{r}(r)$ in terms of the integrated
Gaussian curvature divided by the length of a boundary circle of
radius r, with the result
\begin{equation}
E^r = \frac{1-\sqrt{l(r)}}{r \!\!\!\! \quad l(r)} \!\!\!\! \quad ,
\label{E}
\end{equation}
where we used Equations (\ref{metricappendix1}) and
(\ref{gamma-exp}). The angular component $E^{\phi}$ is zero
everywhere. Note that $E^{r}(r)$ vanishes linearly with $r$ for
small $r$ and decays like $r e^{-\frac{r^{2}}{r_{0}^2}}$ for $r
\gg r_0$. From Eq.(\ref{E}) one can obtain the geometric potential
$V(r)$ by performing a line integral,
\begin{equation}
V(r)=\int_{r}^{\infty}\! dr' \!\!\!\!\! \quad g_{rr} \!\!\!\!\!
\quad E^{r} = - \int_{r}^{\infty} \! dr' \frac{\sqrt{l(r')}-1}{r'}
\!\!\!\! \quad , \label{potential2}
\end{equation}
which matches the result previously obtained in
Eq.(\ref{potential4}).

\section {\label{appC} Free energy on a corrugated plane}

\begin{figure}
\includegraphics[width=0.47\textwidth]{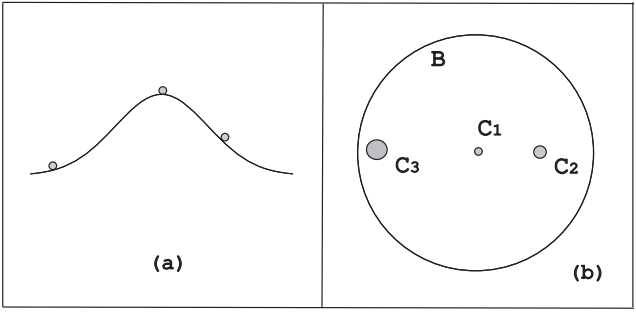}
\caption{\label{fig:app-C} (a) Defects with fixed core size $a$ on
a Gaussian bump encircled by a circular boundary of radius $R$
denoted by $B$. (b) The size of the vortex cores varies with
position when plotted in the conformal plane. One can avoid the
singularities associated with the defects cores by puncturing the
conformal plane. This introduces circular boundaries $C_{i}$ of
varying radius at the position of each defect in the conformal
plane, reflecting corresponding constant core radii on the
Gaussian bump.}
\end{figure}

In this Appendix, we derive the effective free energy for a charge
neutral configuration of defects confined on an infinite surface
of varying Gaussian curvature with the topology of the plane. A
general method was introduced in Ref.\cite{Vite-Turn04} that
allows treatments of the more complicated case of deformed
spheres. A detailed treatment of boundary-effects is developed in
Appendix \ref{appD}. Here we simply assume that the size of the
system is much larger than the size of the bump and that the
boundary does not impose any topological constraint to the
director of the liquid crystal. The results presented here match
those obtained in Appendix \ref{appD} for free boundary
conditions, as long as the defects are far from the boundary.
Suppose that all the $N_d$ defects have the same circular core
radius $a$ which does not depend on where they are located on the
surface. This assumptions is justified if the radius of curvature
$\frac{r_{0}}{\alpha}$ is much greater than $a$. In this limit,
the microscopic physics that determines $a$ is insensitive to the
presence of the curvature and since the bump is locally flat $a$
is approximately constant everywhere. The starting point of our
analysis is the free energy expressed in terms of the singular
part of the bond angle $\theta_{s}({\bf u})$
\begin{equation}
F = \frac{1}{2}K_{A}\int_{S} dA \!\!\!\!\! \quad g^{\alpha\beta}
(\partial_{\alpha}\theta_{s} -
A_{\alpha})(\partial_{\beta}\theta_{s} - A_{\beta})\!\!\!\! \quad
. \label{eq:patic-ener-bis}
\end{equation}
The cores of the defects are excluded from the area integral in
Eq.(\ref{eq:patic-ener-bis}), hence $S$ is a disconnected domain
corresponding to the corrugated surface punctured at the positions
${\bf u}_{i}=(r_{i},\phi_{i})$ of the defects. In the conformal
plane parameterized by the coordinate $\Re(r)$ defined in
Eq.(\ref{solution2}), the defect cores are circles whose position
dependent radius is given by $a e^{V(r_{i})}$. The boundaries of
the "core-disks" are labelled by $C_{i}$ while the circular edge
of the sample of radius $R$ is denoted by $B$, see Fig.
\ref{fig:app-C}.
Upon introducing a "Cauchy conjugate" function $\chi({\bf u})$
defined by
\begin{equation}
\partial_{\alpha}\theta_{s} - A_{\alpha}=\gamma_{\alpha}\!\!\!\!\!\! \quad  ^{\beta} \partial_{\beta} \chi \!\!\!\! \quad ,
\label{eq:chi}
\end{equation}
the free energy in Eq.(\ref{eq:patic-ener-bis}) can be cast in the
form:
\begin{equation}
F = \frac{1}{2}K_{A}\int_{S} dA \!\!\!\!\! \quad g^{\alpha\beta}
(\partial_{\alpha}\chi)(\partial_{\beta}\chi) \!\!\!\! \quad .
\label{eq:patic-ener-bis2}
\end{equation}
In deriving Eq.(\ref{eq:patic-ener-bis2}) we used the identity
\begin{equation}
g^{\mu\nu} \gamma_{\mu} \!\!\!\!\!\! \quad ^{\alpha} \gamma_{\nu}
\!\!\!\!\!\! \quad  ^{\beta}= g^{\alpha\beta} \!\!\!\! \quad ,
\label{ident1}
\end{equation}
which can be proved with the aid of Eq.(\ref{gamma1}) and the
discussion following it. Eq.(\ref{ident1}) implies that the
(covariant) dot product between two vectors after rotating each of
them by $\frac{\pi}{2}$ is equivalent to taking the dot product
between the two initial vectors. The integral in
Eq.(\ref{eq:patic-ener-bis2}) can be rewritten as
\begin{equation}
\frac{F}{K_{A}} = \frac{1}{2}\int_{S} dA \!\!\!\!\! \quad
D_{\alpha}(\chi D^{\alpha} \chi)- \frac{1}{2}\int_{S} dA
\!\!\!\!\! \quad \chi D_{\alpha} D^{\alpha} \chi \!\!\!\!\! \quad
, \label{eq:int-parts}
\end{equation}
where
\begin{equation}
D_{\alpha}(\chi D^{\alpha}\chi)\equiv
(1/\sqrt{g})\partial_{\alpha}(\sqrt{g}
g^{\alpha\beta}\chi\partial_{\beta}\chi )\!\!\!\! \quad ,
\label{eq:def-flux}
\end{equation}
and $D^{\alpha}D_{\alpha}$ is defined in Eq.(\ref{laplacian}).
Upon taking an additional covariant derivative, we can recast
Eq.(\ref{eq:chi}) in the form of a Poisson equation for the
electrostatic-like potential $\chi({\bf u})$,
\begin{equation}
D_{\alpha}D^{\alpha}\chi({\bf u}) = - \rho({\bf u}) \!\!\!\! \quad
, \label{eq:sol-chi}
\end{equation}
where the analogue of the electrostatic charge density $\rho({\bf
u})$ is given by
\begin{equation}
\rho({\bf u}) \equiv \sum_{i=1}^{N_{d}} q_{i} \frac{\delta( {\bf
u},{\bf u}_{i})}{\sqrt{g}}-G({\bf u}) \!\!\!\! \quad . \label{rho}
\end{equation}
It is useful to compare Eq.(\ref{eq:sol-chi}) to
Eq.(\ref{eq-potent}) used to define the geometric potential in
Appendix B. Both expressions are Poisson equations, the only
difference being that the source term of Eq.(\ref{rho}) includes
both the point-like charges of the defects and the Gaussian
curvature with its sign reversed. Hence the Gauss law discussed in
Appendix \ref{appB} for the geometric field $E_{\alpha}$ applies
also to $\partial_{\alpha}\chi$, provided that Eq.(\ref{Gauss}) is
suitably modified to include the contribution from the topological
charges of the defects:
\begin{equation}
\oint_{C} \! d u^{\alpha} \gamma_{\alpha} \!\!\!\!\!\! \quad
^{\beta}
\partial_{\beta} \chi = \int_{A} \! d^{2}u \sqrt{g} \!\!\!\quad
\left(G({\bf u})-\sum_{i=1}^{N_{d}} q_{i} \frac{\delta( {\bf
u},{\bf u}_{i})}{\sqrt{g}}\right) \!\!\!\! \quad , \label{Gauss2}
\end{equation}
where $C$ is the contour enclosing an arbitrary surface $A$. This
relation will be useful later.

We can formally solve for $\chi({\bf u})$ in Eq.(\ref{eq:sol-chi})
in terms of the Green's function $\Gamma({\bf u},{\bf u}')$ found
in Appendix \ref{appA}:
\begin{equation}
\chi({\bf u}) = \int_{A} \! dA' \!\!\!\quad  \rho({\bf u}')
\!\!\!\quad \Gamma({\bf u},{\bf u}') \!\!\!\! \quad ,
\label{eq:sol-chi0}
\end{equation}
where boundary terms have been dropped using charge neutrality and
the fact that the edge of the sample is assumed to be far away
from the defects. The integral in Eq.(\ref{eq:sol-chi0}) can be
evaluated, with the result
\begin{equation}
\chi({\bf u}) = \sum_{i=1}^{N_{d}} q_{i} \Gamma({\bf u},{\bf
u}_{i})-\int_{A} \! dA' \!\!\!\quad  G({\bf u}') \!\!\!\quad
\Gamma({\bf u},{\bf u}') \!\!\!\! \quad . \label{eq:sol-chi1}
\end{equation}
Upon using Eq.(\ref{geometric potential}), we obtain:
\begin{equation}
\chi({\bf u}) = \sum_{i=1}^{N_{d}} q_{i} \Gamma({\bf u},{\bf
u}_{i})+ V({\bf u}) \!\!\!\! \quad . \label{eq:sol-chi2}
\end{equation}
We note that the first term is singular at positions ${\bf u}_{i}$
but when $\chi$ is substituted in Eq.(\ref{eq:patic-ener-bis2})
the resulting energy is finite because the core of the defects are
excluded from the domain of integration $S$. Upon substituting
Eq.(\ref{eq:sol-chi}) in the second term of
Eq.(\ref{eq:int-parts}) we obtain:
\begin{eqnarray}
\int_{S} dA \!\!\!\!\! \quad \chi D_{\alpha} D^{\alpha} \chi &=&
-\int_{S} dA \!\!\!\!\! \quad \chi \!\!\!\!\! \quad \rho({\bf u})\nonumber\\
&=& \int_{S} dA \!\!\!\!\! \quad \chi \!\!\!\!\! \quad G({\bf u})
\!\!\!\! \quad , \label{eq:insert}
\end{eqnarray}
where we dropped terms involving the delta functions in
Eq.(\ref{rho}) because they vanish everywhere except at the
coordinates of the defects which are excluded from the domain of
integration $S$. Upon substituting Eq.(\ref{eq:sol-chi2}) in
Eq.(\ref{eq:insert}) we obtain:
\begin{eqnarray}
&-& \int_{S} dA \!\!\!\!\! \quad \chi D_{\alpha}
D^{\alpha} \chi = \sum_{i=1}^{N_{d}} q_{i} V({\bf u}_i) \nonumber\\
&+& \int dA \!\!\!\!\! \quad \int \! dA' \!\!\!\quad G({\bf u})
\!\!\!\quad \Gamma({\bf u},{\bf u}') \!\!\!\quad G({\bf u}')
\!\!\!\! \quad , \label{eq:IIterm}
\end{eqnarray}
where we used Eq.(\ref{geometric potential}).

To evaluate the first term in Eq.(\ref{eq:int-parts}), we apply
the generalized Stokes formula of Eq.(\ref{Coulomb model2}) and
convert the surface integrals into line integrals over the
boundaries:
\begin{eqnarray}   \int_{S}\! dA \!\!\!\quad
D_{\alpha}(\chi D^{\alpha}\chi) &=&
\sum_{i=1}^{N_{d}}\oint_{C_{i}}\! d u^{\alpha} \chi
\gamma_{\alpha}
\!\!\!\!\!\! \quad ^{\beta} D_{\beta} \chi \nonumber\\
&-& \oint_{B}\! d u^{\alpha} \chi \gamma_{\alpha} \!\!\!\!\!\!
\quad ^{\beta} \chi D_{\beta}\chi \!\!\!\! \quad ,
\label{eq:stok1}
\end{eqnarray}
where the difference in sign between the two boundary integrals in
Eq.(\ref{eq:stok1}) is due to the fact that the outward normals
for the paths $C_i$ are oriented opposite to the normal for $B$,
the outermost boundary of the system.

To evaluate the last term in Eq.(\ref{eq:stok1}), we note that the
flux through the distant boundary $B$ due to a charge neutral
distribution of defects is approximately zero, provided that the
integrated Gaussian curvature enclosed by the boundary is
vanishingly small (see Eq.(\ref{Gauss2})). Hence
\begin{eqnarray}
\chi \oint_{B}\! d u^{\alpha} \gamma_{\alpha} \!\!\!\!\!\! \quad
^{\beta} D_{\beta}\chi \simeq 0 \!\!\!\! \quad , \label{eq:stok3}
\end{eqnarray}
where we used the fact that $\chi$, defined in Eq.(\ref{eq:chi}),
is approximately constant on $B$ since $\partial_{\alpha}\theta -
A_{\alpha}\simeq 0$. By contrast, the flux of $\partial_{r} \chi$
piercing the boundary $C_{i}$ in Eq.(\ref{eq:stok1}) is
approximately equal to the charge $q_{i}$ of the inclosed defect
\footnote{The integrated Gaussian curvature in the microscopic
disk is vanishingly small.}. In evaluating the integrals around
the infinitesimal boundaries $C_{i}$, we used the fact that the
function $\chi({\bf u}_{i}+{\bf a})$ evaluated on the "rim" of the
defect core centered at ${\bf u}_{i}$ and of radius $a e^{V({\bf
u}_{i})}$ is dominated by a logarithmically diverging contribution
due to the $i^{th}$ defect. This leading contribution is
approximately constant on $C_{i}$. On the other hand, the non
diverging part of $\chi({\bf u}_{i}+{\bf a})$ is multiplied by the
perimeter of $C_{i}$ and hence its contribution is of the order of
$a$. The result of the integration will be insensitive to the
orientation of the vectors along the boundary $C_{i}$, provided
the defect core is small \footnote{If the defect core is very
large, it may be necessary to place images within the defect core
itself to impose a desired boundary condition on its rim. This is
not the regime considered in the present work (see Ref.
\cite{Budz94} for similar calculations performed in flat space).}.
In this way, we find
\begin{eqnarray}
\oint_{C_{i}} \! d u^{\alpha} \gamma_{\alpha} \!\!\!\!\!\! \quad
^{\beta} \chi D_{\beta} \chi &\simeq& \chi({\bf u}_{i}+{\bf a})
\oint_{C_{i}} \! d u^{\phi} \gamma_{\phi} \!\!\!\!\!\! \quad
^{r}\partial_{r} \chi \nonumber\\ &\simeq& q_{i} \chi({\bf u}_{i}+
{\bf a}) \!\!\!\! \quad . \label{eq:stok2}
\end{eqnarray}

Upon substituting Equations (\ref{eq:stok2}) and (\ref{eq:stok3})
in Eq.({\ref{eq:stok1}}), we obtain
\begin{eqnarray}
\int_{S} dA \!\!\!\!\! \quad D_{\alpha}(\chi D^{\alpha} \chi) &=&
\sum_{i=1}^{N_{d}}q_{i} \chi({\bf u}_{i}+ {\bf a})= \nonumber\\
\sum_{i=1}^{N_{d}} q_{i} V({\bf u}_i) &+&
\sum_{i=1}^{N_{d}}\sum_{j
\neq i}^{N_{d}} q_{i} q_{j} \Gamma({\bf u}_{i},{\bf u}_{j}) \nonumber\\
&+& \sum_{i=1}^{N_{d}} q_{i}^2 \Gamma({\bf u}_{i},{\bf u}_{i} +
{\bf a}) \!\!\!\! \quad , \label{eq:Iterm}
\end{eqnarray}
where we used Eq.(\ref{eq:sol-chi2}) to substitute for $\chi$.
Substitution of Eq.(\ref{eq:Iterm}) and Eq.(\ref{eq:IIterm}) in
Eq.(\ref{eq:int-parts}) then yields:
\begin{eqnarray}
\frac{F}{K_{A}} &=& F_{0} +
\sum_{i=1}^{N_{d}}q_{i}(1-\frac{q_{i}}{4 \pi})V(r_{i}) - \sum_{i=1}^{N_{d}} \frac{{q_{i}}^{2}}{8 \pi} \ln (a^{2}) \nonumber\\
&+& \frac{1}{2}\sum_{i=1}^{N_{d}} \sum_{j \neq i}^{N_{d}} q_{i}
q_{j} \!\!\!\!\! \quad \Gamma({\bf u}_i,{\bf u}_j) \!\!\!\! \quad
, \label{eq:last}
\end{eqnarray}
where we have used Eq.(\ref{green3}) to evaluate $\Gamma({\bf
u}_i,{\bf u}_i + {\bf a})$. The first term in Eq.(\ref{eq:last})
is the free energy of the smooth defect-free texture (see
Eq.(\ref{free texture energy}) and preceding discussion). The free
energy difference $\frac{{\triangle F}}{K_{A}}$ between a charge
neutral defect configuration and a smooth texture thus reads
\begin{eqnarray}
\frac{\Delta F (\alpha)}{K_{A}} &=& \frac{1}{2}\sum_{i=1}^{N_{d}}
\sum_{j \neq i}^{N_{d}} q_{i} q_{j}
\Gamma_{a}(r_{i},\phi_{i},r_{j},\phi_{j})+ \frac{E_{c}}{K_{A}}\sum_{i=1}^{N_{d}}q_{i}^{2}\nonumber\\
&+& \sum_{i=1}^{N_{d}}q_{i}\left(1-\frac{q_{i}}{4
\pi}\right)V(r_{i}) \!\!\!\! \quad , \label{eq:diff-energy2}
\end{eqnarray}
where the subscript $a$ in the Green's function indicates
\begin{eqnarray} \Gamma_{a}(r_{i},\phi_{i},r_{j},\phi_{j})&=&-\frac{1}{4 \pi} \ln[\frac{\Re(r)^{2}}{a^{2}}+\frac{\Re(r')^{2}}{a^{2}}\nonumber\\
&-& 2\frac{\Re(r)}{a} \frac{\Re(r')}{a} \cos(\phi-\phi')] \!\!\!\!
\quad . \label{green-cut}
\end{eqnarray}
In order to absorb the core radius $a$ in the inter-defect
interaction, we used the elementary algebraic identity
\begin{equation}
\sum_{i=1}^{N_{d}}{q_{i}^{2}}=- \sum_{i=1}^{N_{d}} \sum_{j \neq
i}^{N_{d}} q_{i} q_{j} \!\!\!\! \quad , \label{green2}
\end{equation}
valid for charge neutral configurations. The core energy $E_{c}$
in Eq.(\ref{eq:diff-energy2}) was added by hand and represents
short distance physics on scales less than or equal to the core
radius. Although the second part of the last term in
Eq.(\ref{eq:diff-energy2}) arises as a position dependent self
energy, it has the same functional form as the geometrical
potential discussed in Appendix \ref{appB}, and hence depends on
the global shape of the surface.

\section {\label{appD} Bump with a boundary}

\begin{figure}
\includegraphics[width=0.45\textwidth]{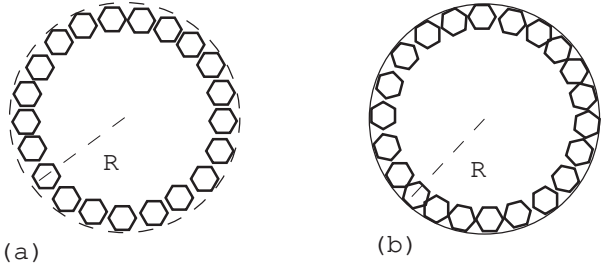}
\caption{\label{fig:appD1}(a) Schematic illustration of the
boundary director texture corresponding to free boundary
conditions. The vector order parameter orientation close to the
edge of the sample does not vary appreciably as one moves along
the radial direction and it is parallel to itself at every point
on the boundary (b) Tangential boundary conditions. The vector
order parameter is locally aligned to a wall located at the edge
of the sample.}
\end{figure}

The aim of this appendix is to study the energetics of a singular
vector field on a bumpy surface of circular shape and finite size
$R$. To evaluate the ground state energy in
Eq.(\ref{eq:patic-ener-bis}), we first need to solve the covariant
Laplace equation for the bond angle field $\theta({\bf u})$ in the
presence of $N_d$ defects at positions ${\bf u}_{i}$. This is more
easily done by switching from $\theta({\bf u})$ to the conjugate
field $\chi({\bf u})$, as shown in Eq.(\ref{eq:patic-ener-bis2}),
and solving the Poisson Equation (\ref{eq:sol-chi}) satisfied by
$\chi$ for both free and fixed boundary conditions (see Fig.
\ref{fig:appD1}).

The bond angle field satisfies free boundary conditions if the
following relation holds on the circular edge $B$:
\begin{equation}
\partial_{r} \theta |_{r=R} = 0 \!\!\!\! \quad , \label{bc-free}
\end{equation}
while for fixed tangential boundary conditions we have:
\begin{equation}
\partial_{\phi} \theta |_{r=R} = 0 \!\!\!\! \quad . \label{bc-circ}
\end{equation}
To understand Eq.(\ref{bc-circ}), recall that we measure the bond
angle $\theta$ with respect to a rotating basis vector $E_{r}$ in
the radial direction. With this convention, $\theta$ is equal to a
constant when the vector order parameter is aligned with the
circular boundary $B$. We can convert Equations (\ref{bc-free})
and (\ref{bc-circ}) into boundary conditions to be satisfied by
the conjugate field $\chi$ on $B$. Upon substituting
Eq.(\ref{bc-free}) in Eq.(\ref{eq:chi}) and using the fact that
$A_{r}$ is equal to zero, we obtain the constraint that $\chi({\bf
u})$ satisfies on $B$ in the case of free boundary conditions:
\begin{equation}
\partial_{\phi} \chi |_{r=R} = 0 \!\!\!\! \quad . \label{bc-free2}
\end{equation}
This corresponds to a Dirichlet problem where $\chi^{D}({\bf u})$
evaluated on the boundary $B$ assumes an arbitrary constant value,
$c$:
\begin{equation}
\chi^{D}(B) = c \!\!\!\! \quad . \label{bc-free2bis}
\end{equation}

Upon substituting Eq.(\ref{bc-circ}) in Eq.(\ref{eq:chi}), we
obtain
\begin{equation}
\partial_{r} \chi = -\frac{A_{\phi}}{\gamma_{\phi} \!\!\!\!\!\! \quad
^{r}} \!\!\!\! \quad , \label{bc-circ-int}
\end{equation}
since $\gamma_{\phi}^{\phi}=0$. Upon substituting Equations
(\ref{gamma-exp}) and (\ref{A_phi}) in Eq.(\ref{bc-circ-int}) we
obtain the boundary condition on $\chi$ that corresponds to
Eq.(\ref{bc-circ})
\begin{equation}
\partial_{r} \chi^{N} |_{r=R} = -\frac{1}{R} \!\!\!\! \quad , \label{bc-circ2}
\end{equation}
where the superscript indicates that this is a Neumann boundary
problem with the normal derivative assuming a constant value.

To solve the Poisson Eq.(\ref{eq:sol-chi}) with Neumann or
Dirichlet's boundary conditions in terms of suitable Green's
functions we exploit a covariant version of Green's theorem
expressed in terms of two invariant functions of position
$\psi({\bf u})$ and $\varphi({\bf u})$ \cite{McConnell-book}:
\begin{eqnarray}
\int_{S} \! dA  \!\!\!\!\quad \left(\varphi({\bf u}) \!\!\!\!\!
\quad D_{\alpha} D^{\alpha} \psi({\bf u})-\psi({\bf u}) \!\!\!\!\!
\quad D_{\alpha} D^{\alpha} \varphi({\bf u})\right) = \nonumber\\
- \oint_{B} \! d u^{\alpha} \gamma_{\alpha} \!\!\!\!\!\! \quad
^{\beta} \left(\varphi({\bf u}) \!\!\!\!\! \quad \partial_{\beta}
\!\!\!\!\! \quad \psi({\bf u})-\psi({\bf u}) \!\!\!\!\! \quad
\partial_{\beta} \!\!\!\!\! \quad \varphi({\bf u})\right)
 \!\!\!\! \quad .
\label{eq:chi-bound}
\end{eqnarray}
By applying Eq.(\ref{eq:chi-bound}) to $\varphi({\bf u})=\chi({\bf
u})$ and $\psi({\bf u})=\Gamma ({\bf u},{\bf u'})$ and using
Equations (\ref{poisson}) and (\ref{eq:sol-chi}) we obtain
\begin{eqnarray}
\chi({\bf u})&=& \int_{S} \! dA ' \!\!\!\quad \Gamma({\bf u'},{\bf
u}) \!\!\!\quad \rho({\bf u'}) \nonumber\\ &+& \oint_{B} \! d
u'^{\alpha} \gamma_{\alpha} \!\!\!\!\!\! \quad ^{\beta} \chi({\bf
u'}) \!\!\!\!\! \quad
\partial'_{\beta} \Gamma ({\bf u'},{\bf u}) \nonumber\\ &-& \oint_{B} \! d
u'^{\alpha} \gamma_{\alpha} \!\!\!\!\!\! \quad ^{\beta}\Gamma
({\bf u'},{\bf u}) \!\!\!\!\! \quad
\partial'_{\beta}\chi({\bf u'})
 \!\!\!\! \quad ,
\label{eq:chi-bound2}
\end{eqnarray}
where ${\bf u}$ and ${\bf u'}$ have been exchanged \footnote{These
manipulations are common in electromagnetism, see for example Ref.
\cite{Wyld-book}.}. The boundary conditions for the Green's
function can be conveniently chosen to eliminate unknown
quantities in Eq.(\ref{eq:chi-bound}) as in flat space
\cite{Wyld-book}.

For the Dirichlet's problem, we choose the Green's function
$\Gamma^{D}$ so that it vanishes when ${\bf u}'$ is on the
boundary $B$
\begin{equation}
\Gamma^{D} (B, {\bf u}) = 0 \!\!\!\! \quad . \label{gf1}
\end{equation}
Upon substituting Eq.(\ref{gf1}) in Eq.(\ref{eq:chi-bound2}) and
noting that $\chi({\bf u'})$ is constant on the boundary $B$ (see
Eq.(\ref{bc-free2bis})), we obtain:
\begin{eqnarray}
\chi^{D}({\bf u})&=&\int_{S} \! dA ' \!\!\!\quad \Gamma ^{D}({\bf
u'},{\bf u}) \!\!\!\quad \rho({\bf u'}) \nonumber\\ &+&
\chi^{D}(B) \oint_{B} \! d u'^{\alpha} \gamma_{\alpha}
\!\!\!\!\!\! \quad ^{\beta} \!\!\!\!\! \quad
\partial'_{\beta} \Gamma ^{D} ({\bf u'},{\bf u})
 \!\!\!\! \quad ,
\label{eq:chi-bound3}
\end{eqnarray}
The contour integral in the second term of
Eq.(\ref{eq:chi-bound3}) corresponds to the flux piercing $B$
which is, in turn, equal to the (unit) charge of the singularity
located at ${\bf u}$ \footnote{This assertion can be proved by
applying Stokes Theorem (as stated in Eq.(\ref{Coulomb model2})
with $E_{\beta}$ replaced by $\partial'_{\beta} \Gamma ({\bf
u'},{\bf u})$) and using Eq.(\ref{poisson})) to evaluate the
surface integral.}. The final result reads
\begin{equation}
\chi^{D}({\bf u}) = \int_{S} \! dA' \!\!\!\quad \rho({\bf u}')
\!\!\!\quad \Gamma^{D}({\bf u'},{\bf u}) + \chi^{D}(B) \!\!\!\!
\quad , \label{eq:sol-chi0bis}
\end{equation}
where $\chi^{D}(B)$ can be set to zero, since the energy in
Eq.(\ref{eq:patic-ener-bis2}) is only defined in terms of
derivatives of $\chi$. One can check that $\chi^{D}({\bf u})$ in
Eq.(\ref{eq:sol-chi0bis}) satisfies both the Poisson's Equation
(\ref{eq:sol-chi}) and the required boundary condition in
Eq.(\ref{bc-free2}). This can be more easily proved by noting that
$\Gamma^{D}({\bf u'},{\bf u})$ is symmetric under exchange of its
arguments ${\bf u'}$ and ${\bf u}$ \footnote{This assertion can be
proved by applying Green's theorem in Eq.\ref{eq:chi-bound} to
$\psi({\bf u})=\Gamma ({\bf u},{\bf u'})$ and $\varphi({\bf
u})=\Gamma ({\bf u},{\bf u ''})$ and noticing that the right hand
side vanishes if the boundary condition in Eq.(\ref{gf1}) is
assumed. We can then conclude that $\Gamma^{D}({\bf u'},{\bf
u''})=\Gamma^{D}({\bf u''},{\bf u'})$ in analogy with familiar
results in 3D electrostatics \cite{Wyld-book}.}.

A similar reasoning applies to $\chi^{N}({\bf u})$. However, we
cannot choose the Green's function so that the second term of
Eq.(\ref{eq:chi-bound2}) (that contains the unknown quantity
$\chi({\bf u'})$) vanishes. In fact, by invoking Stokes theorem
(see Eq.(\ref{Coulomb model2})), we note that
\begin{equation}
\oint_{B} \! d u'^{\alpha} \gamma_{\alpha} \!\!\!\!\!\! \quad
^{\beta} \!\!\!\!\! \quad
\partial  \!\!\!\!\! \quad ' _{\beta} \Gamma ({\bf u'},{\bf u})=1 \!\!\!\!\!
\quad , \label{int-1}
\end{equation}
where $\gamma_{\alpha} \!\!\!\!\!\! \quad ^{\beta}({\bf u'}=B)$ is
constant on the circular boundary B and can be brought out of the
integral. An appropriate choice of boundary condition on
$\Gamma^{N}$ that satisfies the constraint in Eq.(\ref{int-1}) is
\begin{eqnarray}
\partial_{r'} \!\!\!\!\! \quad \Gamma^{N}(r',\phi ';r,\phi)| _{r'=R} &=& \frac{1}{2 \pi \gamma_{\alpha} \!\!\!\!\!\! \quad ^{\beta}(R)} \nonumber\\ &=& - \frac{\sqrt{l(R)}}{2 \pi R} \!\!\!\! \quad ,
\label{gf2}
\end{eqnarray}
where we used Eq.(\ref{gamma-exp}). The $\alpha$-dependent
function $l(r')$ was defined in Eq.(\ref{l-bis}). Note that $l(R)
\simeq 1$, for $R \gg r_{0}$ (see Fig. \ref{fig:l(r)}). By
substituting Equations (\ref{gf2}) and (\ref{bc-circ2}) in
Eq.(\ref{eq:chi-bound2}) we obtain
\begin{eqnarray}
\chi^{N}({\bf u})&=&\int_{S} \! dA ' \!\!\!\quad \Gamma ^{N}({\bf
u'},{\bf u}) \!\!\!\quad \rho({\bf u'}) \nonumber\\ &+& \frac{1}{2
\pi}\int_{0}^{2 \pi} d \phi '  \!\!\!\!\! \quad \chi(R,\phi ')
\nonumber\\ &-& \frac{1}{\sqrt{l(R)}} \int_{0}^{2 \pi} d \phi '
\!\!\!\!\! \quad \Gamma^{N}(R,\phi ';r,\phi)
 \!\!\!\! \quad ,
\label{eq:chi-bound4}
\end{eqnarray}
The last two integral are constant and hence can be dropped. We
can check explicitly that $\chi^{N}({\bf u})$ satisfies
Eq.(\ref{bc-circ2}) by evaluating the radial derivative of
$\chi^{N}({\bf u})$ in Eq.(\ref{eq:chi-bound4})
\begin{equation}
\partial_{r}\chi^{N}(r,\phi)| _{r=R} = \int_{S} \! dA' \!\!\!\quad \rho({\bf u}')
\!\!\!\quad \partial_{r} \Gamma^{N}(R, \phi';r,\phi) | _{r=R}
\!\!\!\! \quad , \label{eq:proof}
\end{equation}
where the radial derivative of the Green's function assumes the
constant value derived in Eq.(\ref{gf2}), provided that
$\Gamma^{N}({\bf u}',{\bf u})$ is constructed so that it is
symmetric under exchange of its arguments ${\bf u}'$ and ${\bf u}$
(see Eq.(\ref{eq:wideeq})). With the aid of Equations (\ref{rho})
and (\ref{intgr Gauss curv}), we obtain
\begin{eqnarray}
\int_{S} \! dA' \!\!\!\quad \rho({\bf u}') = \sum_{i}^{N_{d}}
q_{i} +2 \pi \left( \frac{1}{\sqrt{l(R)}}-1 \right) \!\!\!\! \quad
. \label{cond1}
\end{eqnarray}
Upon substituting Equations (\ref{cond1}) and (\ref{gf2}) in
Eq.(\ref{eq:proof}), we conclude that the the Neumann boundary
condition in Eq.(\ref{bc-circ2}) is satisfied provided that
\begin{equation}
\sum_{i}^{N_{d}} q_{i} = 2 \pi \!\!\!\! \quad . \label{cond2}
\end{equation}
It is reassuring that the topological constraint on the vorticity
of the field imposed by the presence of the wall arises as a
natural requirement within this formalism. Similarly, the
Poisson's equation for $\chi^{N}({\bf u})$ is automatically
satisfied.

\begin{figure}
\includegraphics[width=0.45\textwidth]{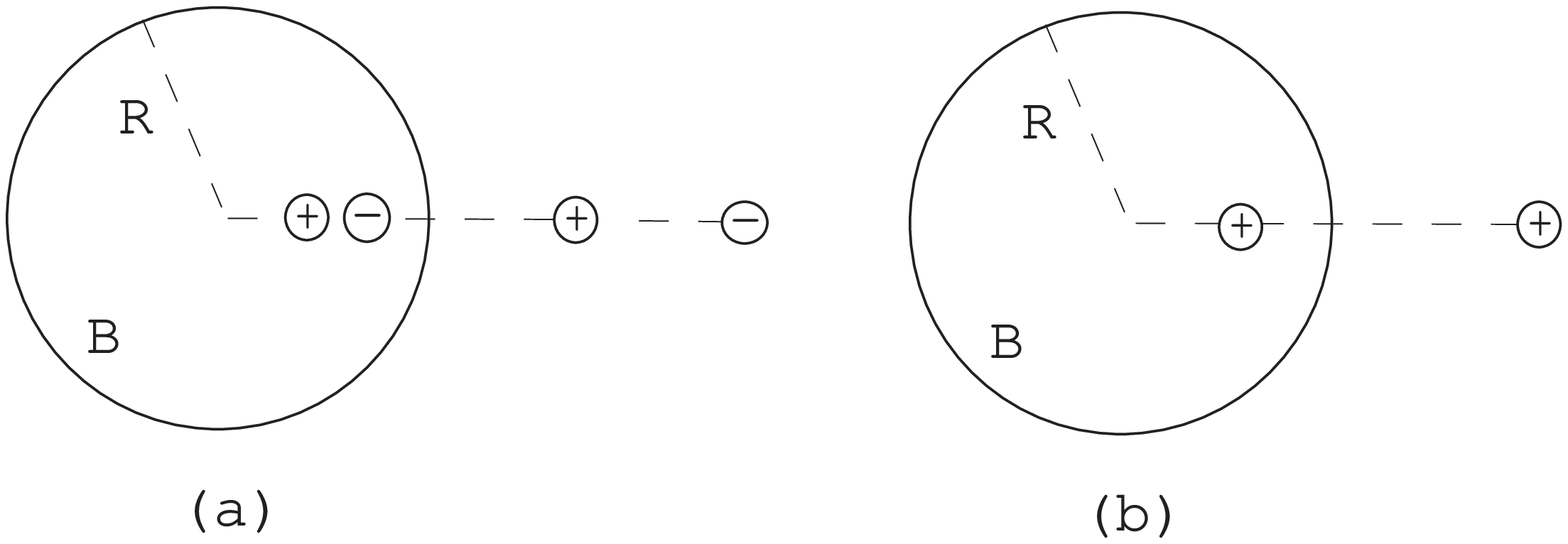}
\caption{\label{fig:appD2}Schematic illustration of the method of
images. The image defect is of the same sign for free boundary
conditions (a) and opposite for fixed boundary conditions (b).
Defects closer to the center of the circle have images further
away from it.}
\end{figure}

\begin{figure}
\includegraphics[width=0.40\textwidth]{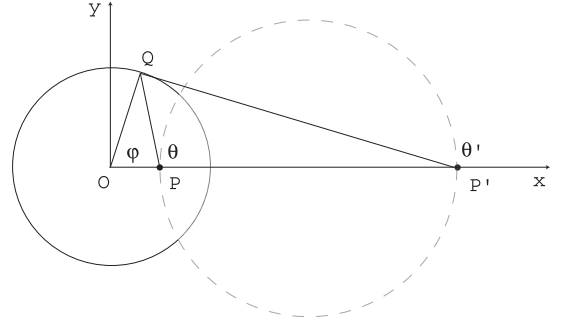}
\caption{\label{fig:boundary}A topological defect located at
position $P$ in a circular domain of radius $\overline{OQ}$ in
flat space. Fixed boundary conditions are obtained by placing an
image defect of the same sign at a distance $\overline{OP'}$ from
the center such that $\overline{OP}$ $\overline{OP'}$ $=$
$\overline{OQ}^{2}$. The two triangles $\bigtriangleup OQP$ and
$\bigtriangleup OQP'$ are similar and $\angle OQP = \angle OP'Q =
\pi - \theta'$. By the theorem of the external angle, we conclude
that $\theta ' + \theta = \phi + \pi$ as long as $Q$ lies on the
circumference $B$. This is equivalent to the boundary condition in
Eq.(\ref{bc-circ}) if a non rotating vector basis is used.
Similarly, for free boundary conditions the symmetric Green'
function evaluated is constant on the boundary if the image defect
is negative. Since
$\frac{\overline{PQ}}{\overline{P'Q}}=\frac{\overline{OP}}{\overline{OQ}}$,
the potential $ \ln ( \overline{PQ}) - \ln ( \overline{P'Q})$
(generated by the defect at distance $\overline{OP}$ and its
image) is constant on the circumference of radius
$\overline{OQ}$.}
\end{figure}

We are now left with the task of guessing the Green's functions
for the Dirichlet's and Neumann problems satisfying the boundary
conditions in Equations (\ref{gf1}) and (\ref{gf2}) respectively.
In both cases the Green's function can be determined by the method
of images applied in the conformal plane. For every defect with
radial coordinate $r_{i}$ we need an image defect of opposite
(equal) topological charge at position $r'_{i}$ to ensure that
Dirichlet's (Neumann) boundary conditions are enforced (see Fig.
\ref{fig:appD2}).
The radial coordinate of the
image defect $r'_{i}$ is determined by the relation:
\begin{equation}
\Re(r'_{i}) = \frac{\Re \!\!\!\!\! \quad ^{2}(R)}{\Re(r_{i})}
\!\!\!\! \quad . \label{pos-ima}
\end{equation}
Except for the coordinate change $r \rightarrow \Re(r)$, a similar
relation arises in elementary electrostatic problems in flat space
\cite{Pano-book}. A geometric argument that justifies this choice
of images in flat space is illustrated in Fig. \ref{fig:boundary}.

Once the position of the source is chosen according to
Eq.(\ref{pos-ima}), we can express the two Green's functions with
the concise notation $\Gamma^{D/N}$ as follows:
\begin{widetext}
\begin{equation}
\Gamma^{D/N}({\bf u},{\bf u'})=- \frac{1}{4 \pi} \ln
[\Re(r)^{2}+\Re(r')^{2}-2\Re(r)\Re(r')\cos (\phi - \phi ')] \pm
\frac{1}{4 \pi} \ln[\Re(r)^{2}+\frac{\Re(R)^{4}}{\Re(r')^2} -
2\Re(r)\frac{{\Re(R)}^{2}}{\Re(r')}\cos (\phi - \phi ')] \pm
f(r')\;, \label{eq:wideeq}
\end{equation}
\end{widetext}
where we have introduced a function $f(r')$ to make
$\Gamma^{D/N}({\bf u},{\bf u'})$ symmetric under exchange of its
arguments and to remove a singularity at $r'=0$. Note that we can
add $f(r')$ since the defining equation of the Green's function
does not contain derivatives of $r'$, only of $r$.
\begin{equation}
f(r') = \frac{1}{2 \pi}\ln\left[\frac{\Re \!\!\!\!\! \quad
(r')}{\Re(R)}\right] \!\!\!\! \quad . \label{f-s}
\end{equation}
The plus and minus signs in Eq.(\ref{eq:wideeq}) insure that the
Dirichlet and Neumann boundary conditions respectively are obeyed.
In what follows the sign placed above in the symbols $\pm$ or
$\mp$ always indicates the choice suitable for the Dirichlet's
problem while the one below refers to Neumann's boundary
conditions. One can explicitly check by substitution that the
symmetrized Green's functions $\Gamma^{D/N}({\bf u},{\bf u'})$
satisfy the correct boundary conditions, as long as the plus sign
is chosen when $\Gamma^{D}$ is substituted in Eq.(\ref{gf1}) and
the minus sign when $\Gamma^{N}$ is substituted in Eq.(\ref{gf2}).
Note that, without the extra term $f(r')$ in the expressions for
both Green's functions, $\Gamma^{D}$ would not be equal to zero on
the boundary $B$ and the last term in Eq.(\ref{eq:chi-bound4})
would not be constant when $\Gamma^{N}$ is substituted in.

Once the Green's function is obtained, one can readily write down
$\chi^{D/N}({\bf u})$ by dropping the constant terms in Equations
(\ref{eq:sol-chi0bis}) and (\ref{eq:chi-bound4})
\begin{eqnarray}
\chi^{D/N}({\bf u}) &=& \sum_{i=1}^{N_{d}} q_{i} \Gamma^{D/N}({\bf
u},{\bf u}_{i}) \nonumber\\ &-& \int_{S} \! dA' \!\!\!\quad
G({\bf u}') \!\!\!\quad \Gamma^{D/N}({\bf u},{\bf u}') \!\!\!\!
\quad . \label{eq:sol-chi1D}
\end{eqnarray}
The Gaussian curvature is given by the covariant Laplacian of the
geometric potential introduced in Eq.(\ref{eq-potent}). Upon
integrating by parts twice the second term in
Eq.(\ref{eq:sol-chi1D}) and applying Stokes theorem repeatedly, we
find
\begin{equation}
\chi^{D/N}({\bf u}) = \sum_{i=1}^{N_{d}} q_{i} \Gamma^{D/N}({\bf
u},{\bf u}_{i})+V({\bf u}) \!\!\!\! \quad , \label{eq:chiD}
\end{equation}
where we assume that $R\gg r_0$ so that we can neglect boundary
terms. The geometric potential $V({\bf u})$ has the same
functional form previously discussed in Appendix \ref{appB},
despite the change in the Green's function.

The evaluation of the energy stored in the field now proceeds
along the lines sketched in Appendix \ref{appC} with the only
caveat that one needs to choose the appropriate Green's function
in Eq.(\ref{eq:wideeq}). In the case of Dirichlet's boundary
conditions one can prove that the boundary integral in
Eq.(\ref{eq:stok3}) still vanishes by virtue of the fact that
$\chi$ is constant on the boundary $B$ and the defects
configuration is charge neutral. For the Neumann's problem we
have:
\begin{eqnarray}
\chi^{N} \oint_{B}\! d u^{\alpha} \gamma_{\alpha} \!\!\!\!\!\!
\quad ^{\beta} D_{\beta}\chi^{N} \approx 4\pi \ln\left( \Re(R)
\right) \!\!\!\! \quad , \label{eq:stokD}
\end{eqnarray}
where we assumed that $R \gg r_{0}$. In this limit $\Re(R)$ is
approximately equal to $R$, as can be checked with the aid of
Equations (\ref{solution2}) and (\ref{eq:pot-bis}).

All the remaining intermediate steps to derive the free energy
follow as in Appendix \ref{appC} without further assumptions. We
can readily generalize Eq.(\ref{eq:diff-energy2}) to evaluate the
energy stored in the singular field in the presence of a boundary
in the case of both free and fixed boundary conditions. We assume
$R \gg r_{0}$, but the defects do not need to be far away from the
boundary. The result is
\begin{eqnarray}
\frac{F^{D/N}}{K_{A}} &=& \frac{1}{2}\sum_{j \neq i}^{N_{d}} q_{i}
q_{j} \!\!\!\!\! \quad \Gamma^{D/N}(x_{i};x_{j}) + F_{0}
\nonumber\\&+& \sum_{i=1}^{N_{d}}q_{i}(1-\frac{q_{i}}{4
\pi})V(r_{i})+ \sum_{i=1}^{N_{d}} \frac{{q_{i}}^{2}}{4 \pi} \ln
\left[\frac{\Re(R)}{a}\right]\nonumber\\&\pm& \sum_{i=1}^{N_{d}}
\frac{{q_{i}}^{2}}{4 \pi} \ln \left[1-x_{i}^{2}\right] \!\!\!\!
\quad , \label{eq:longD}
\end{eqnarray}
where $F_{0}$ is defined in Eq.(\ref{free texture energy}). The
Green's function expressed in scaled coordinates reads
\begin{eqnarray}
\Gamma^{D/N}(x_{i};x_{j})=-\frac{1}{4 \pi} \ln
\left[x_{i}^{2}+x_{j}^{2}-2x_{i} x_{j} \cos \left( \phi_{i}-
\phi_{j}\right) \right]\nonumber\\ \pm \frac{1}{4 \pi} \ln
\left[x_{i}^{2} x_{j}^{2} + 1 - 2 x_{i} x_{j} \cos \left(\phi_{i}-
\phi_{j}\right) \right] \!\!\!\! \quad . \nonumber\\
\label{eq:green-norm}
\end{eqnarray}
In the case of Neumann's boundary conditions, we have suppressed a
term diverging like $\ln(\Re(R))$ associated with the boundary
contribution in Eq.(\ref{eq:stokD}). Eq.(\ref{eq:green-norm}) is
expressed in terms of a dimensionless defect position $x_{i}$
\begin{eqnarray}
x_{i} \equiv \frac{\Re(r_{i})}{\Re(R)} \!\!\!\! \quad .
\label{eq:scaled-coord}
\end{eqnarray}
The plus sign in Equations (\ref{eq:green-norm}) and
(\ref{eq:longD}) is to be chosen for Dirichlet's boundary
conditions and the minus sign for Neumann's. The last term in
Eq.(\ref{eq:longD}) represents the interaction ,
$U_{b}^{D/N}(x_{i})$, between a single defect located at $x_{i}$
and the boundary
\begin{eqnarray}
U_{b}^{D/N}(x_{i})= \pm K_{A}  \sum_{i=1}^{N_{d}}
\frac{{q_{i}}^{2}}{4 \pi} \ln \left[1-x_{i}^{2}\right] \!\!\!\!
\quad . \label{eq:uwall}
\end{eqnarray}
Note that the q-dependent prefactors of $U_{b}^{D/N}(x_{i})$ and
the quadratic correction to the curvature interaction (III term in
Eq.(\ref{eq:longD}) have the same magnitude. This is not a
coincidence but a clue to their common origin. As the geometry of
a plane is modified, either by creating a varying curvature or
imposing boundaries, the defects feel an additional interaction
caused by the conformal transformation of the underlying space.
This line of reasoning is powerful and it has been pursued in Ref.
\cite{Vite-Turn04} to explain some basic features of the
interaction between defects and curvature without explicit
recourse to the Green's function techniques adopted in this work.

\bibliography{bump_mod}

\end{document}